\documentclass[iop]{emulateapj}

\usepackage{natbib}
\usepackage{amsmath, amssymb}
\usepackage{times}
\usepackage{apjfonts, graphicx, graphics}
\usepackage[breaklinks,colorlinks,urlcolor=blue,citecolor=blue]{hyperref}
\usepackage[all]{hypcap}
\usepackage{aas_macros}

\newcommand{\Mpc}{\rm\; Mpc}
\newcommand{\kpc}{\rm\; kpc}
\newcommand{\pc}{\rm\; pc}
\newcommand{\km}{\rm\; km}
\newcommand{\m}{\rm\; m}
\newcommand{\cm}{\rm\; cm}
\newcommand{\mm}{\rm\; mm}

\newcommand{\cmpssq}{\hbox{$\cm\s^{-2}\,$}}

\newcommand{\yr}{\rm\; yr}
\newcommand{\Gyr}{\rm\; Gyr}
\newcommand{\Myr}{\rm\; Myr}
\newcommand{\s}{\rm\; s}

\newcommand{\GHz}{\rm\; GHz}
\newcommand{\MHz}{\rm\; MHz}

\newcommand{\K}{\rm\; K}

\newcommand{\Msun}{\hbox{$\rm\thinspace M_{\odot}$}}

\newcommand{\Msunpyr}{\hbox{$\Msun\yr^{-1}\,$}}

\newcommand{\keV}{\rm\; keV}

\newcommand{\erg}{\rm\; erg}
\newcommand{\Jy}{\rm\; Jy}
\newcommand{\mJy}{\rm\; mJy}

\newcommand{\ergps}{\hbox{$\erg\s^{-1}\,$}}

\newcommand{\kmps}{\hbox{$\km\s^{-1}\,$}}

\newcommand{\kmpspMpc}{\hbox{$\kmps\Mpc^{-1}\,$}}

\newcommand{\Lsun}{\hbox{$\rm\thinspace L_{\odot}$}}

\newcommand{\Zsun}{\hbox{$\thinspace \mathrm{Z}_{\odot}$}}

\newcommand{\pcmsq}{\hbox{$\cm^{-2}\,$}}

\providecommand{\e}[1]{\ensuremath{\times 10^{#1}}}

\newcommand{\uJy}{\rm\; \mu Jy}
\newcommand{\beam}{\rm\; beam}
\newcommand{\Jykmps}{\hbox{$\Jy\km\s^{-1}\,$}}

\newcommand{\uJypbeam}{\hbox{$\uJy\beam^{-1}\,$}}

\shorttitle{Molecular Gas in Zw3146}
\shortauthors{Vantyghem et al.}

\submitted{February 2, 2021}

\begin{document}

\title{A Massive, Clumpy Molecular Gas Distribution and Displaced AGN in Zw~3146}
\author{A.~N. Vantyghem$^{1}$}
\author{B.~R. McNamara$^{2,3,4}$}
\author{C.~P. O'Dea$^{1}$}
\author{S.~A. Baum$^{1}$}
\author{F. Combes$^{5, 6}$}
\author{A.~C. Edge$^{7}$}
\author{A.~C. Fabian$^{8}$}
\author{M. McDonald$^{9}$}
\author{P.~E.~J. Nulsen$^{9, 10}$}
\author{H.~R. Russell$^{8}$}
\author{P. Salom{\'e}$^{5}$}

\affil{
	$^1$ University of Manitoba, Department of Physics and Astronomy, Winnipeg, MB R3T 2N2, Canada; \href{mailto:adrian.vantyghem@umanitoba.ca}{adrian.vantyghem@umanitoba.ca} \\
    $^2$ Department of Physics and Astronomy, University of Waterloo, Waterloo, ON N2L 3G1, Canada \\ 
    $^3$ Waterloo Centre for Astrophysics, Waterloo, ON N2L 3G1, Canada \\
    $^4$ Perimeter Institute for Theoretical Physics, Waterloo, Canada \\
    $^5$ LERMA, Observatoire de Paris, PSL University, CNRS, Sorbonne University Paris, France \\
    $^6$ Coll{\`e}ge de France, 11 place Marcelin Berthelot, 75005 Paris \\
    $^7$ Department of Physics, Durham University, Durham DH1 3LE \\
    $^8$ Institute of Astronomy, Madingley Road, Cambridge CB3 0HA \\
    $^9$ Harvard-Smithsonian Center for Astrophysics, 60 Garden Street, Cambridge, MA 02138, USA \\
    $^{10}$ ICRAR, University of Western Australia, 35 Stirling Hwy, Crawley, WA 6009, Australia \\    
}

\begin{abstract}
We present a recent ALMA observation of the CO(1-0) line emission in the central galaxy of the Zw~3146 galaxy cluster ($z=0.2906$). We also present updated X-ray cavity measurements from archival {\it Chandra} observations. The $5\times 10^{10}\,M_{\odot}$ supply of molecular gas, which is confined to the central 4~kpc, is marginally resolved into three extensions that are reminiscent of the filaments observed in similar systems. No velocity structure that would be indicative of ordered motion is observed. The three molecular extensions all trail X-ray cavities, and are potentially formed from the condensation of intracluster gas lifted in the wakes of the rising bubbles. Many cycles of feedback would be require to account for the entire molecular gas reservoir. The molecular gas and continuum source are mutually offset by 2.6~kpc, with no detected line emission coincident with the continuum source. It is the molecular gas, not the continuum source, that lies at the gravitational center of the brightest cluster galaxy. As the brightest cluster galaxy contains possible tidal features, the displaced continuum source may correspond to the nucleus of a merging galaxy. We also discuss the possibility that a gravitational wave recoil following a black hole merger may account for the displacement.
\end{abstract}

\keywords{
    galaxies: active --- 
    galaxies: clusters: individual (ZwCl 3146) --- 
    galaxies: ISM --- 
    galaxies: kinematics and dynamics
}

\section{Introduction}

Galaxy clusters are permeated by a hot ($10^7-10^8\K$), diffuse intracluster medium (ICM) that shines brightly in X-rays. In cool core galaxy clusters the central density is sharply peaked at the position of the brightest cluster galaxy (BCG). The correspondingly high X-ray luminosity in the cluster core should result in the ICM condensing into clouds of cold, molecular gas and forming stars at rates of hundreds to thousands of solar masses per year \citep{Fabian94, Peterson06}.
Indeed, the BCGs in cool core clusters are the largest and most luminous elliptical galaxies in the Universe. They are burgeoning with star formation proceeding at rates of $\sim 1-500\Msunpyr$ \citep[e.g.][]{mcn04, ODea08, McDonald11, Donahue15, Tremblay15, McDonald18}, host luminous emission-line nebulae \citep[e.g.][]{Lynds70, Heckman81, Cowie83, Hu85, Crawford99}, and harbour molecular gas reservoirs more massive than even gas-rich spiral galaxies, sometimes exceeding $10^{10}\Msun$ \citep{Edge01, Edge02, Edge03, Salome03}. 
These signatures of gas condensation are prevalent in systems whose central cooling times fall below a sharp threshold of $1\Gyr$ \citep{Cavagnolo08, Rafferty08, Pulido18}. Nevertheless, they account for less than 10\% of what would be expected from an unimpeded cooling flow.

Active galactic nuclei (AGN) feedback regulates the rate of ICM cooling \citep[for reviews, see][]{McNamara07, McNamara12, Fabian12}. The ``radio-mode'' of AGN feedback is nearly ubiquitous in giant ellipticals and galaxy clusters with short central cooling times \citep{Burns90, Dunn06, Best07, Birzan12}. Radio jets launched by the central AGN inflate bubbles, drive shock fronts, and generate sound waves in the hot atmosphere \citep[e.g.][]{mcn00, Churazov00, Churazov01, Blanton01, Fabian06}. The sizes and surrounding pressures of radio bubbles, which manifest as cavities in the X-ray surface brightness, are used in a direct estimate of the AGN power output. In a large sample of objects, the AGN power output closely matches the radiative losses from the surrounding ICM \citep{Birzan04, Dunn06, Rafferty06, Nulsen09}. This, along with the preponderance of systems with central cooling times $<1\Gyr$, implies that AGN heating and radiative cooling are coupled in a long-lived feedback loop.

In order to establish and maintain such a feedback loop, radiative losses in the ICM must be tied to accretion onto the central supermassive black hole (SMBH). Molecular gas is the most likely intermediary. Overdensities in the ICM, generated by turbulence or AGN feedback, rapidly condense into a multiphase atmosphere that is dominated by molecular gas \citep[e.g.][]{Sharma12, McCourt12, Gaspari13, Gaspari17a, Li14a, Voit18}. Ram pressure and collisions between cold clouds are both effective mechanisms for driving cold clouds toward the galactic center \citep{Pizzolato05, Pizzolato10}. This results in ``chaotic cold accretion'' \citep{Gaspari13}, where the stochastic production of cold clouds feeds black hole growth and regulates the feedback loop between ICM cooling and AGN heating.

Though molecular gas is the fuel for AGN feedback, its distribution is also shaped by the AGN. Radio jets drive fast outflows of ionized, neutral, and molecular gas on $\pc$ to $\kpc$ scales \citep[e.g.][]{Morganti05, Morganti13, Nesvadba06, Alatalo11, Dasyra11, Sturm11, Tadhunter14, Morganti15}.
A gentler coupling between AGN feedback and molecular gas is observed in BCGs \citep{Olivares19, Russell19}. The rotationally-supported disks that would be expected from the long-lived accumulation of molecular clouds in the galactic center are rare \citep{Hamer14, Russell19}.
Instead, billions of solar masses of cold gas are situated in kpc-scale filaments that trail X-ray cavities \citep[e.g.][]{Salome06, Salome08, Lim08, Lim12, McDonald12, mcn14, Russell14, Russell16, Russell17, Russell17b, Vantyghem16, Vantyghem18, Vantyghem19}.
The detection of redshifted absorption lines indicates that some of this gas rains back onto the central galaxy in a circulation flow \citep{David14, Tremblay16, Tremblay18, Rose19a, Rose19b, Rose20}.

Motivated by these observations, \citet{mcn16} proposed the ``stimulated feedback'' paradigm \citep[see also][]{Revaz08}. As radio bubbles rise buoyantly, they uplift low-entropy ICM from the cluster core to an altitude where it becomes thermally unstable. The cold clouds decouple from the radio bubbles and fall back onto the central galaxy as in chaotic cold accretion. In this way, AGN feedback stimulates the production of its own fuel. 

Mapping the spatial distribution of cold gas in BCGs in all but the closest and most gas-rich systems has only become possible with the arrival of the Atacama Large Millimeter Array (ALMA).
Here we present a new ALMA observation targeting the CO(1-0) emission line in the BCG of Zw~3146, which is one of the most extreme cool core clusters and brightest CO emitters known. 
If left unimpeded, its soft ($0.1-2.4\keV$) X-ray luminosity of $2\e{45}\ergps$ \citep{Ebeling98, Bohringer00} would lead to a mass deposition rate of $1250\Msunpyr$ \citep{Edge94}. Indeed, the BCG in Zw~3146 is exceptionally active, showing a blue continuum and luminous nebular emission lines \citep{Allen92, Hicks05}, an infrared luminosity of $4\e{11}\Lsun$ \citep{Egami06}, an $8\e{10}\Msun$ reservoir of cold molecular gas \citep{Edge01, Edge03}, and strong molecular H$_2$ emission lines from $\sim10^{10}\Msun$ of warm H$_2$ \citep{Egami06b}. The infrared luminosity implies a star formation rate of $\sim 70\Msunpyr$ \citep{Egami06, Edge10, Hicks10, Hoffer12}.

Spectroscopic measurements indicate that the true mass deposition rate is much lower than implied by the X-ray luminosity \citep[e.g.][]{Egami06}. Any continuous radiative cooling flow from the bulk temperature at $4\keV$ to $0.01\keV$ is less than $50\Msunpyr$ (Liu et al. in preparation, see also \citealt{Liu19}). The rate of gas cooling to $0.7\keV$ is $130\Msunpyr$, with less than $78\Msunpyr$ cooling to lower temperatures.
AGN feedback is likely responsible for the heavily suppressed cooling rate. The X-ray atmosphere hosts several cavities that replenish energetic losses at a rate of $\sim6\e{45}\ergps$ \citep{Rafferty06}.
The combination of a massive reservoir of molecular gas and powerful AGN feedback make Zw~3146 a prime candidate to contain molecular filaments trailing X-ray cavities, as observed in other BCGs.

Throughout this work we assume a standard $\Lambda$-CDM cosmology with $H_0=70\kmpspMpc$, $\Omega_{{\rm m}, 0}=0.3$, and $\Omega_{\Lambda, 0}=0.7$. At the redshift of Zw~3146 ($z=0.29007$), the angular scale is $1''=4.35\kpc$ and the luminosity distance is $1500\Mpc$.


\section{Observations and Data Reduction}

The BCG of the ZwCl~3146 galaxy cluster (2MASX J10233960+0411116 -- R.A.: 10:23:39.6340, decl.: +4:11:10.660) was observed with ALMA in Band 3 (Cycle 4, ID 2018.1.01056.S, PI Vantyghem), centered on the CO(1-0) line at $89.316\GHz$. The observations were conducted in three blocks from 15-19 August 2019, with a total on-source integration time of 106 minutes. Each observing block was split into $\sim6$ minute on-source integrations interspersed with observations of the phase calibrator. The 69.2 arcsec primary beam was centered on the BCG nucleus in a single pointing common to each observing block. 41-44 antennas were used in these observations, with baselines ranging from $41$ to $3637\m$. This corresponds to a maximum recoverable scale of $10''$. A single spectral window in the frequency division correlator mode was used to study the CO spectral line with 1.875 GHz bandwidth and 488 kHz ($1.6\kmps$) frequency resolution, though the data were later smoothed to a coarser velocity grid. An additional three basebands with the time division correlator mode, each with a 2~GHz bandwidth and frequency resolution of $15.625\MHz$, were employed in order to measure the continuum source.

The observations were calibrated in {\sc casa} version 5.4.0 \citep{casa} using the pipeline reduction scripts. Continuum-subtracted data cubes were created using {\sc uvcontsub} and {\sc clean}. Additional phase self-calibration was attempted, but the continuum source was too faint to provide an improvement in signal-to-noise. Images of the line emission were reconstructed using Briggs weighting with a robust parameter of 0.5. The final CO(1-0) data cube had a synthesized beam of $0.30\times0.26$~arcsec (P.A. $-79.5^{\circ}$) and was binned to a velocity resolution of $20\kmps$. The RMS noise in line-free channels was $2.4\mJy~{\rm beam}^{-1}$.

\subsection{Systemic Velocity}

\begin{figure}
	\includegraphics[width=\columnwidth]{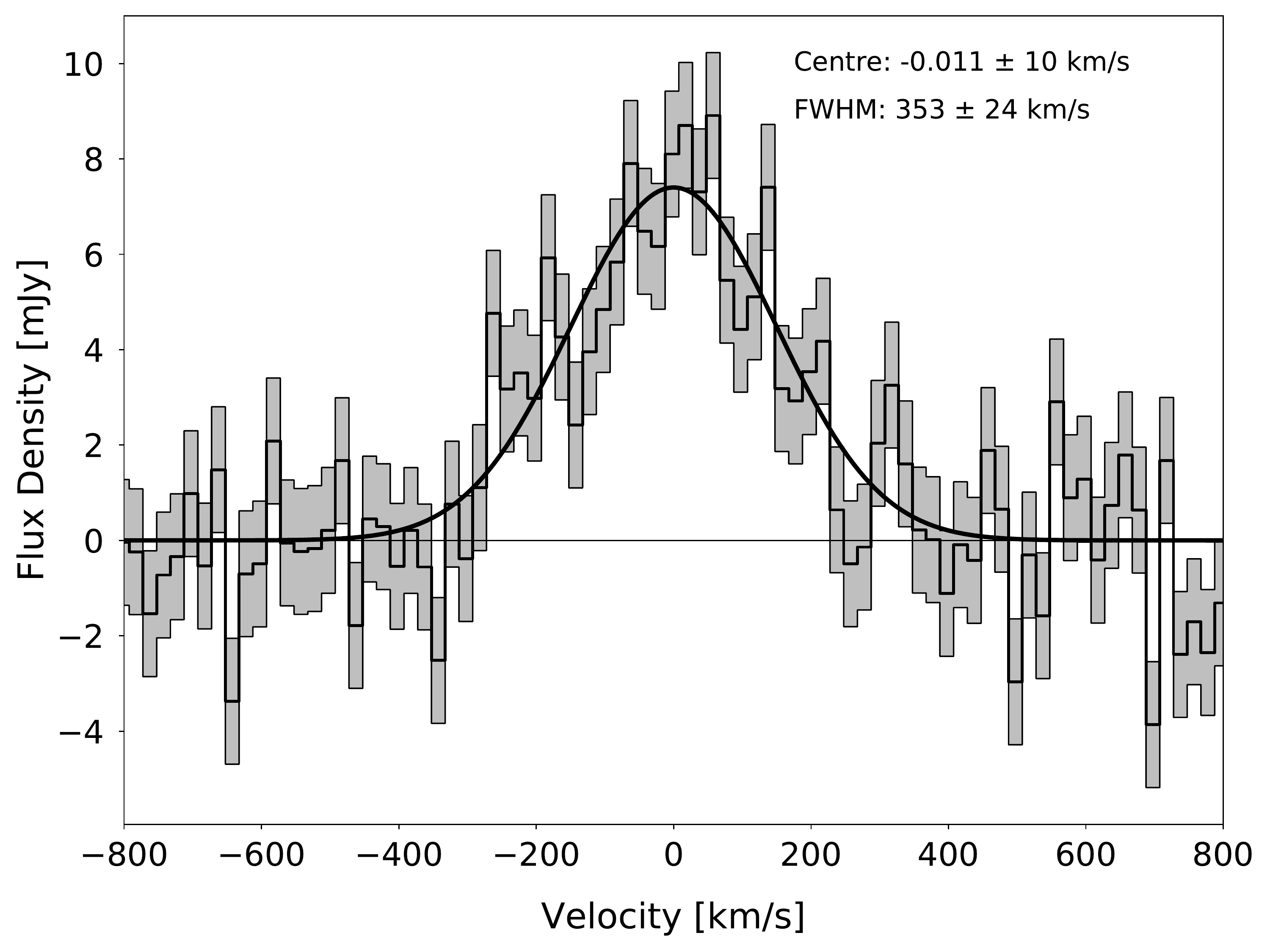}
	\caption{CO(1-0) spectrum extracted from a $13\times 13\kpc$ ($3''\times 3''$) box that contains all of the observed emission.}
	\label{fig:spectrum}
\end{figure}

The gas velocities presented in this work are measured in the molecular gas rest frame. This was determined by extracting a spectrum from a $3''\times 3''$ region that encompasses all of the observed line emission, as shown in Fig. \ref{fig:spectrum}. A single-gaussian fitted to the spectrum yields a molecular gas redshift of $0.29007\pm0.00003$. This is consistent with the CO(1-0) velocity centroids measured in previous work \citep[e.g.][]{Edge01}.

Previous works on Zw~3146 generally adopted a redshift of $0.29060\pm0.00010$, which was determined from the emission-line nebula \citep{Allen92}. The molecular gas rest frame is blueshifted by $158\pm30\kmps$ from this nebular redshift. The discrepancy may originate from differences in the distribution of gas traced by each observation. The molecular gas is confined to the central few kpc of the BCG, while nebular emission is present on larger scales.

\begin{table*}
	\begin{minipage}{\textwidth}
		\caption{Parameters of Molecular Features}
		\begin{center}
			\begin{tabular}{l c c c c c}
				\hline \hline
				Region & $\chi^2/$dof & Velocity Center & FWHM       & Integrated Flux$^{\dagger}$ & Gas Mass$^{\dagger}$ \\
				&              & ($\kmps$)       & ($\kmps$)  & ($\Jykmps$)          & ($10^8\Msun$) \\
				\hline
				All Emission & 221.2/197 & $0.0\pm10$ & $353\pm24$ & $2.78\pm0.17$ & $504\pm30$ \\
				Central Clump & 226.3/197 & $-59.6\pm8.2$ & $265\pm19$ & $0.201\pm0.013$ & $36.4\pm2.3$ \\
				Clump 1 & 224.0/197 & $17.2\pm6.1$ & $177\pm14$ & $0.289\pm0.02$ & $52.7\pm3.7$ \\
				Clump 2 & 221.6/197 & $28\pm12$ & $288\pm28$ & $0.429\pm0.036$ & $78.3\pm6.7$ \\
				Clump 3 & 218.6/197 & $-69\pm17$ & $415\pm41$ & $0.46\pm0.04$ & $83.7\pm7.1$ \\
				\hline \hline
			\end{tabular}
		\end{center}
		Notes: All spectra have been corrected for the response of the primary beam and instrumental broadening. \newline
		\label{tab:spectra}
	\end{minipage}
\end{table*}

\section{Results}

\subsection{Molecular Gas Mass}
\label{sec:Mmol}

The integrated flux ($S_{\rm CO}\Delta v$) of the CO(1-0) line can be converted to molecular gas mass through \citep{Solomon87, Solomon05, Bolatto13}
\begin{equation}
M_{\rm mol} = 1.05\e{4} \frac{X_{\rm CO}}{X_{\rm CO, gal}} 
\left(\frac{S_{\rm CO}\Delta v~D_L^2}{1+z}\right)\Msun,
\label{eqn:Mmol}
\end{equation}
where $z$ is the redshift of the source, $D_L$ is the luminosity distance in Mpc, and $S_{\rm CO}\Delta v$ is in $\Jykmps$. The CO-to-H$_2$ conversion factor, $X_{\rm CO}$, has been studied extensively in the local Universe. 
Its nominal value is the mean empirically-derived Galactic value, $X_{\rm CO, gal} = 2\e{20}\pcmsq (\K\kmps)^{-1}$ \citep{Bolatto13}.

Lacking an independent calibration of $X_{\rm CO}$ in BCGs, we follow standard practice by adopting the Galactic value throughout this work. The only estimate of $X_{\rm CO}$ in a BCG was obtained recently through the detection of $^{13}$CO(3-2) emission in RXJ0821.0+0752 \citep{Vantyghem17}. As $^{13}$CO is an optically thin emission line, it provided a direct measurement of the $^{13}$CO column density. In combination with a measurement of CO(1-0) and CO(3-2) line intensities and a number of assumptions, $X_{\rm CO}$ was estimated to be half of the Galactic value. This is reassuringly close to the Galactic value given the significant morphological differences between BCGs and the Milky Way disk. Until a direct calibration in BCGs is available, we continue to follow the standard practice of adopting the Galactic $X_{\rm CO}$.

The primary reasons for the CO-to-H$_2$ conversion factor to vary between galaxies are the gas metal abundance and excitation conditions. 
In metal-poor dwarf galaxies $X_{\rm CO}$ is elevated by more than an order of magnitude because CO traces a smaller fraction of the overall H$_2$ distribution \citep{Bolatto13}. The molecular gas in BCGs likely forms by condensation of the hot atmosphere, which has typical metallicities of $\sim0.6-0.8\Zsun$ \citep[e.g.][]{DeGrandi01, DeGrandi04}. The metal abundance of molecular clouds in BCGs is therefore expected to be similar to that of the Milky Way.

$X_{\rm CO}$ is driven downward by factors of $\sim5$ in starbursts (e.g. Ultra-Luminous Infrared Galaxies; ULIRGs), where the hot, dense molecular gas has an elevated luminosity per unit mass. BCGs also have high star formation rates, approaching $1000\Msunpyr$ in the most extreme cases \citep[e.g. the Phoenix Cluster;][]{McDonald12Phoenix}. However, the molecular gas in BCGs is often distributed over much larger spatial scales than in ULIRGs, where the gas is confined to the central kpc. Despite their elevated SFRs, BCGs still lie on the Kennicutt-Schmidt relation \citep{Kennicutt98, Kennicutt12} alongside normal galaxies \citep[see e.g.][]{mcn14, Russell14, Vantyghem18}.

The SFR in Zw~3146, $70\Msunpyr$ \citep{Egami06}, is comparable to other BCGs for which the Galactic $X_{\rm CO}$ has been adopted. However, Zw~3146 is also the most luminous and distant H$_2$ emitter observed, with a line luminosity of $L[{\rm H}_2\ 0-0\ S(3)]=6.1\e{42}\ergps$ \citep{Egami06}. A cosmic-ray dominated region or excitation through shocks driven by the AGN may be required to explain such strong rotational H$_2$ lines \citep{Ferland09, Bayet10, Guillard12}. This could impact the value of $X_{\rm CO}$. Strong H$_2$ line emission is observed in many cool core clusters \citep{Donahue11} and radio galaxies \citep{Ogle10}.

A spatially-integrated CO(1-0) spectrum was extracted from a $13\times 13\kpc$ ($3''\times 3''$) box containing all of the detected line emission. The spectrum, shown in Fig. \ref{fig:spectrum}, is best-fit by a single Gaussian with total flux $2.78\pm0.17\Jykmps$. This corresponds to $5.05\pm0.50\e{10}\Msun$ of molecular gas. A list of all fitted parameters is provided in Table \ref{tab:spectra}. No other CO(1-0) line emission was detected throughout the $70''$ primary beam of these observations.

Previous single-dish and interferometric radio observations of Zw~3146 with the IRAM-30m telescope and Owens Valley Radio Telescope (OVRO) yielded CO(1-0) integrated fluxes of $5.2\pm1.2\Jykmps$ and $5.7\pm0.9\Jykmps$, respectively \citep{Edge01, Edge03}. Our ALMA observation recovers 50\% of this flux; a similar fraction to other BCGs \citep[e.g.][]{David14, Russell14, mcn14, Vantyghem18}.
The line emission in the OVRO observations, which recovers all of the single-dish flux, is unresolved by the $6.7'' \times 4.7''$ beam. This suggests that our ALMA observations are missing large-scale flux, predominantly on scales of $\sim5''$ ($22\kpc$).

\begin{figure*}
	\begin{minipage}{0.6325\textwidth}
		\includegraphics[width=\textwidth]{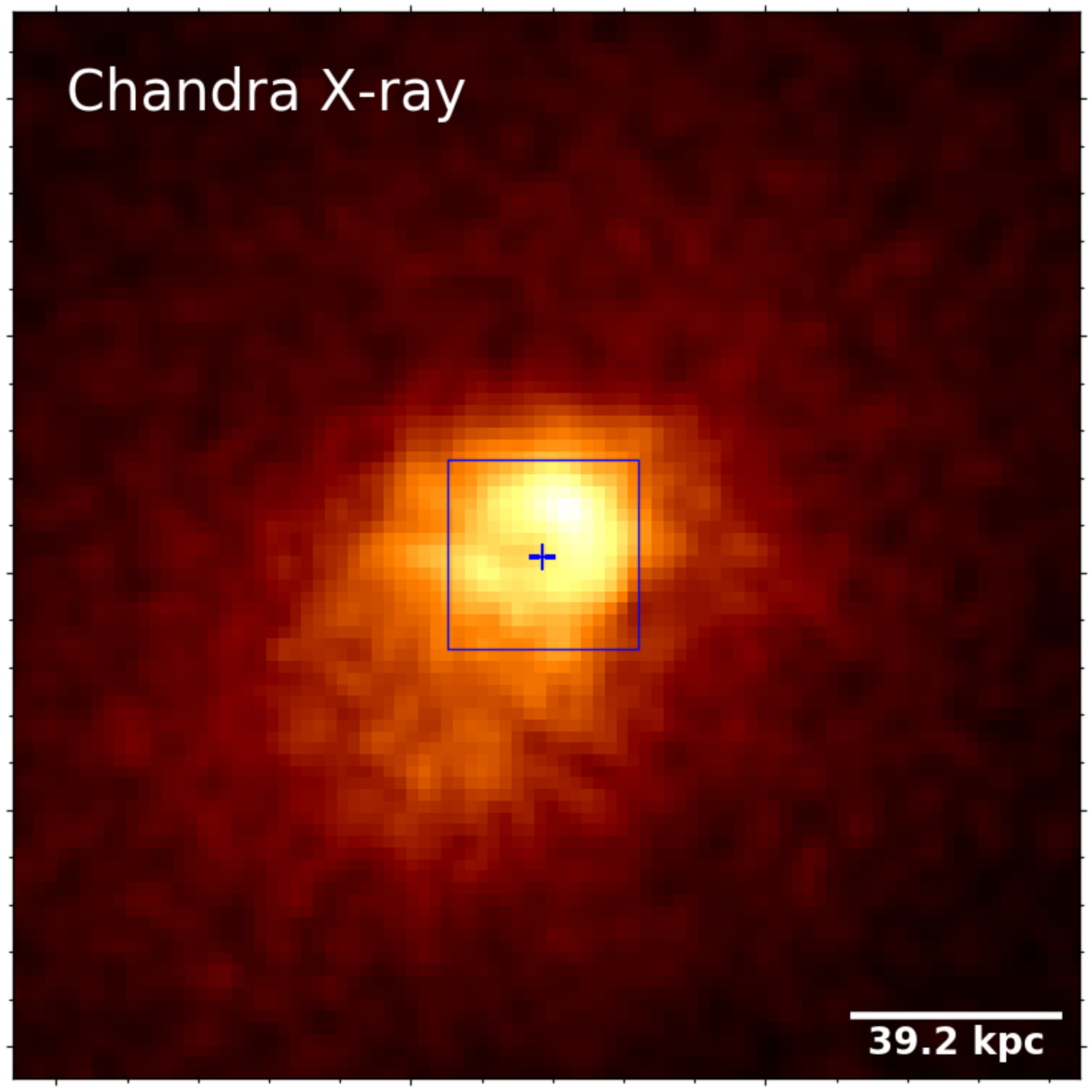}
		\includegraphics[width=\textwidth]{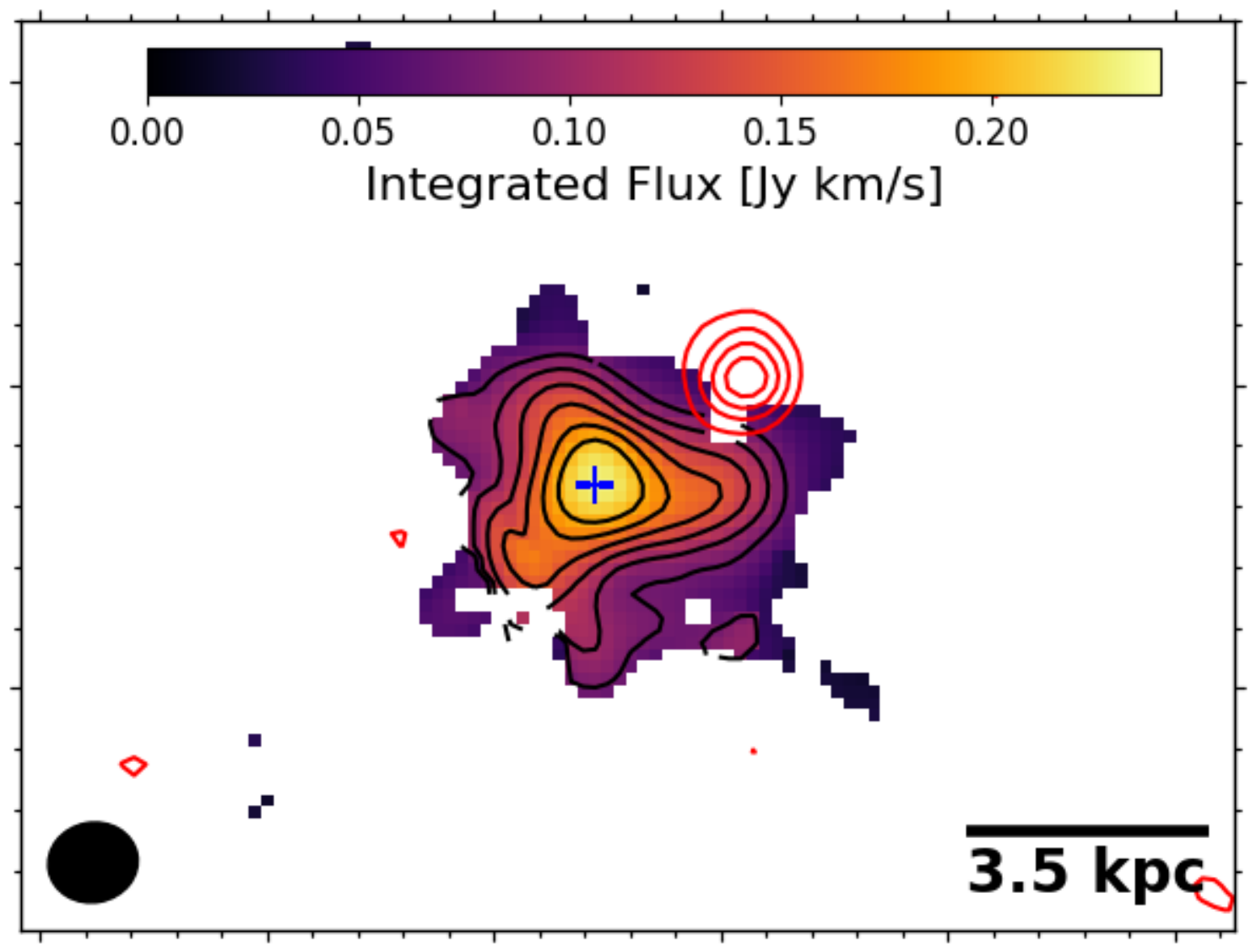}
	\end{minipage}%
	\begin{minipage}{0.3675\textwidth}
		\includegraphics[width=\textwidth]{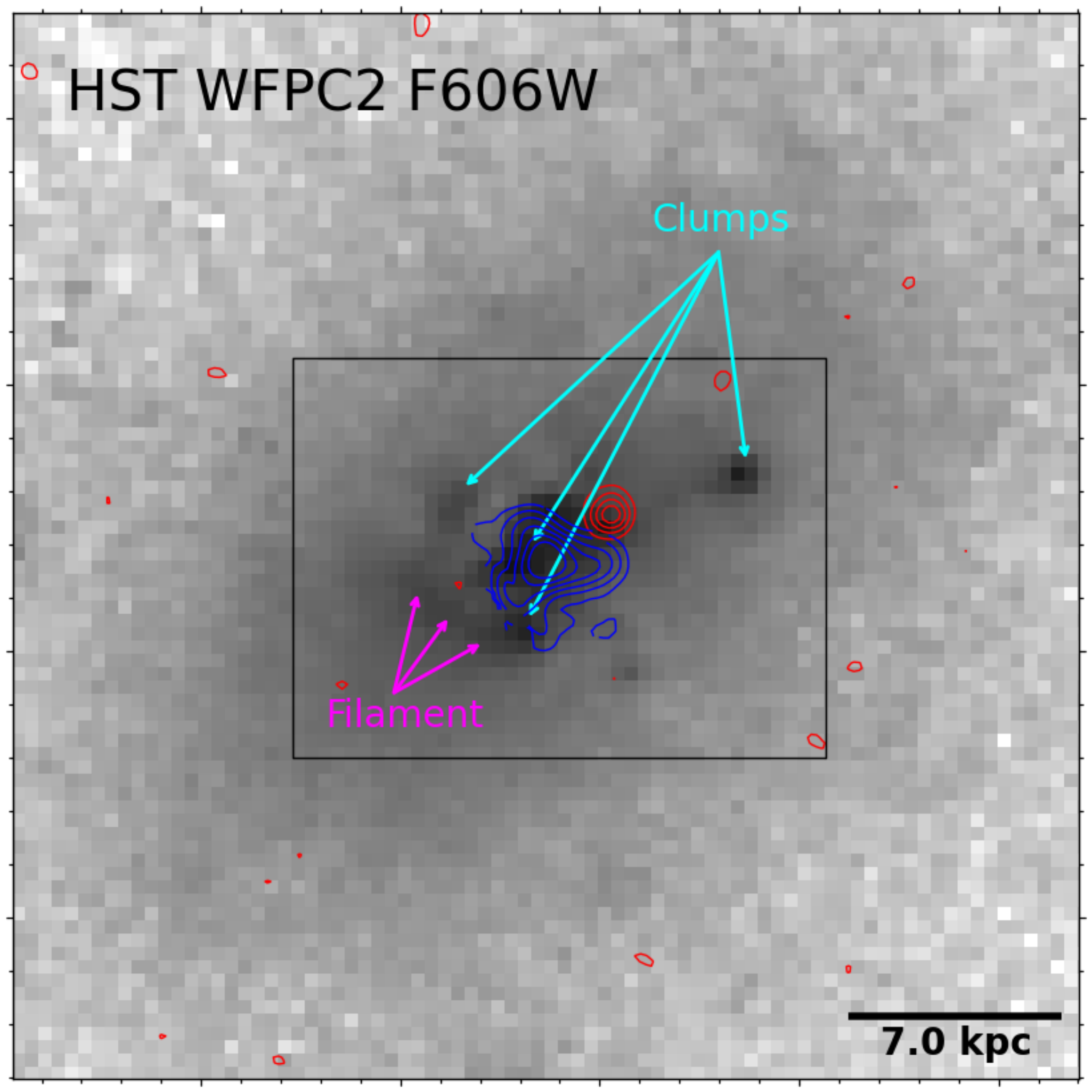}
		\includegraphics[width=\textwidth]{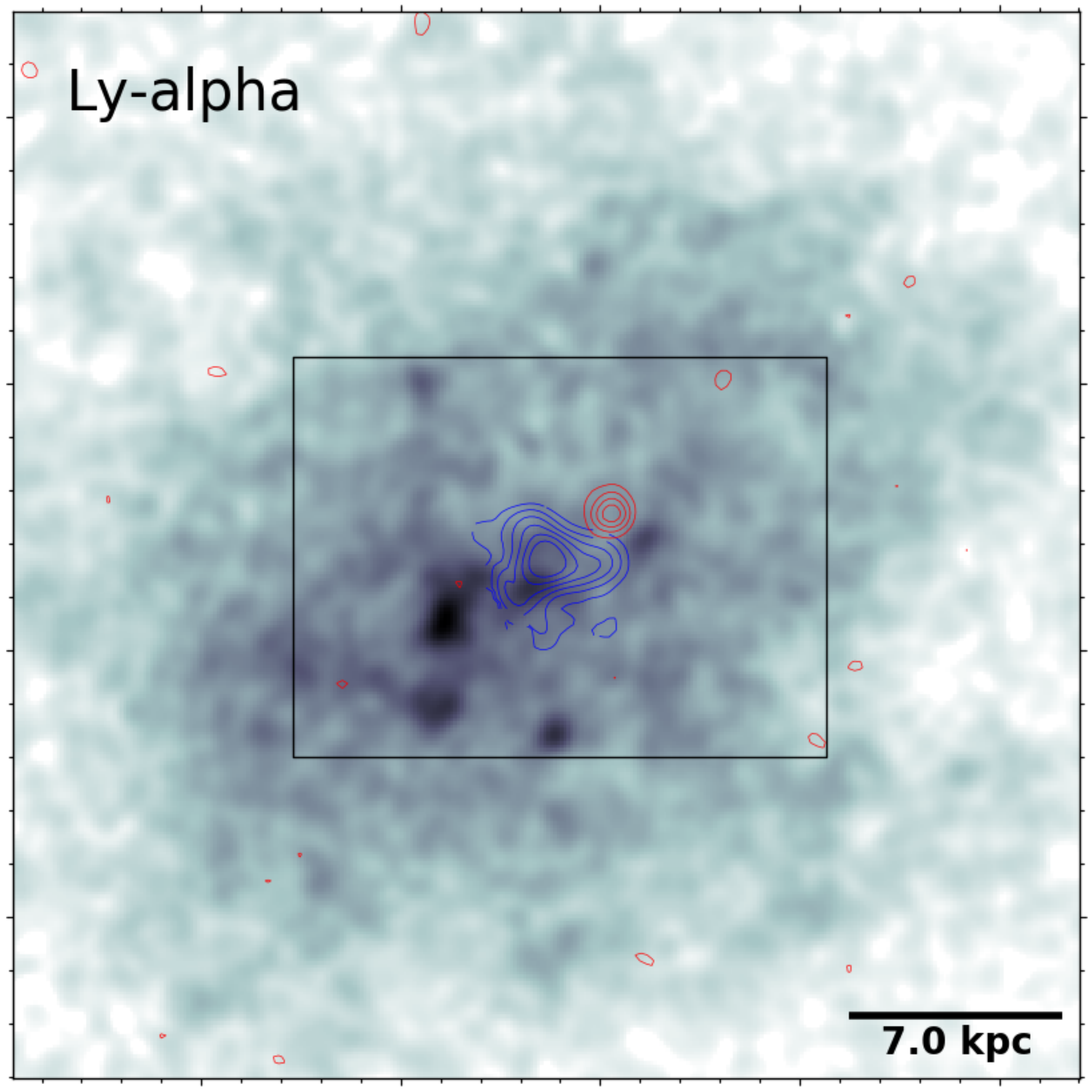}
		\includegraphics[width=\textwidth]{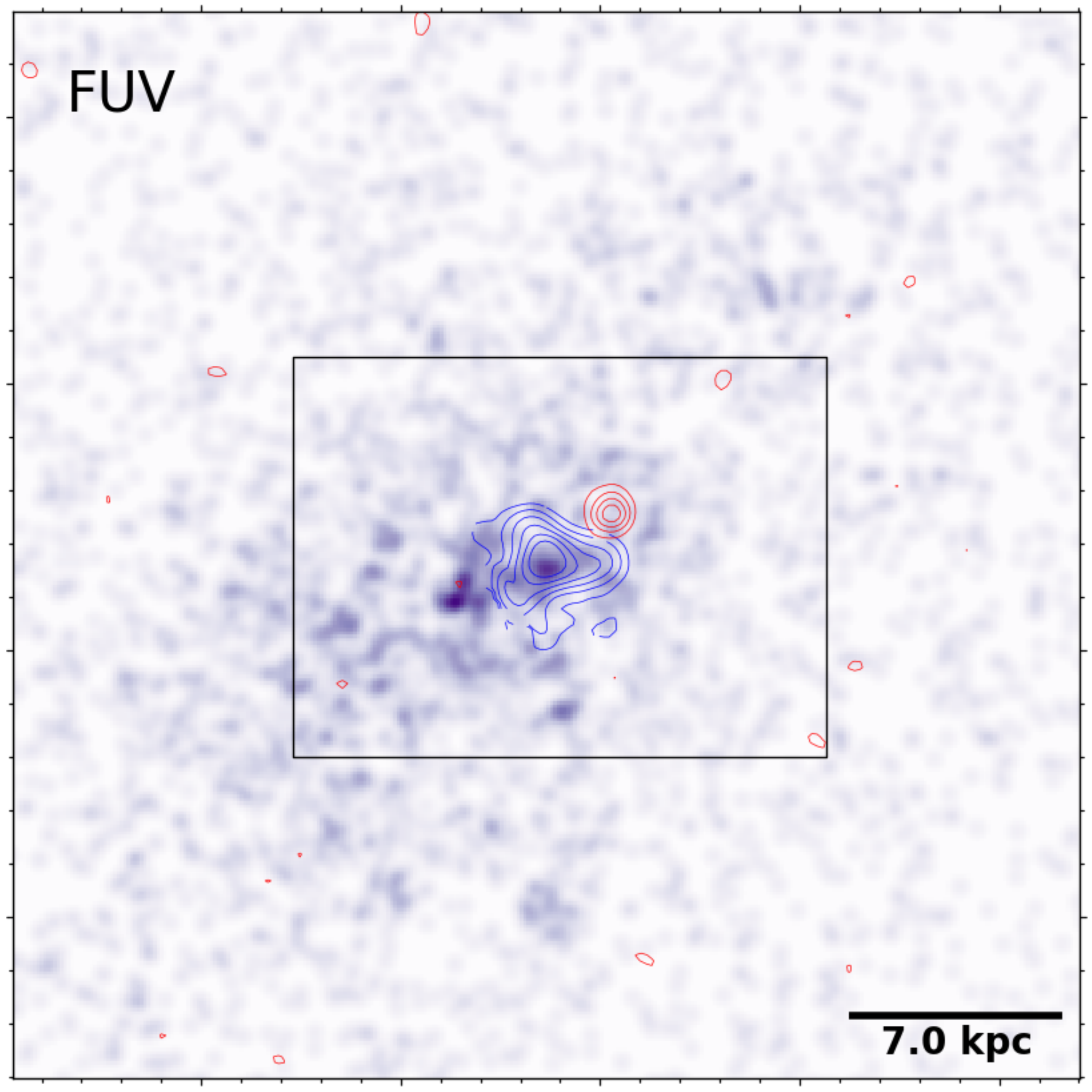}
	\end{minipage}
	\caption{A multiwavelength view of the Zw~3146 cluster. The top-left panel shows the lightly-smoothed {\it Chandra} X-ray $0.5-7\keV$ image. The box indicates the field-of-view for all three images on the right. From top to bottom these are the {\it HST} WFPC2 F606W image, Ly-$\alpha$, and far-UV \citep{Tremblay15}. The most prominent substructures in the HST F606W image have been identified. Note that the F606W filter includes a contribution from H$\beta$ and [O\textsc{iii}] line emission. The rectangle in each of the panels on the right indicate the field-of-view of the ALMA CO(1-0) image, shown on the bottom-left. The solid black ellipse is the synthesized beam. The blue contours and $+$ indicate the CO(1-0) emission and its centroid, respectively. The red contours correspond to $4\sigma$ steps in the $\mm$ continuum, beginning at $3\sigma$ and with $\sigma=7.45\uJypbeam$.}
	\label{fig:comparison}
\end{figure*}

\subsection{Molecular Gas Distribution and Kinematics}
\label{sec:kinematics}

\begin{figure}[t]
	\centering
	\includegraphics[width=0.49\textwidth]{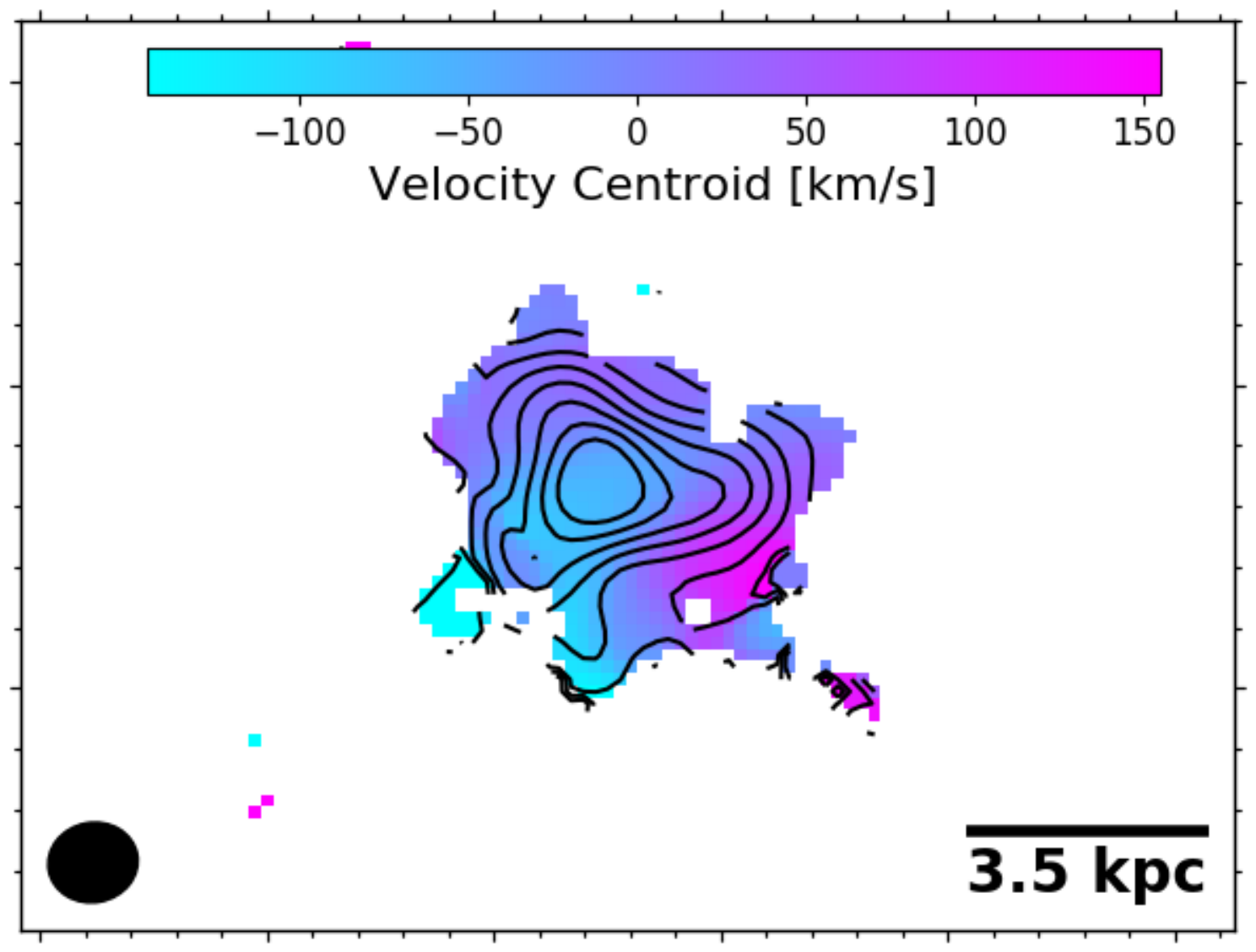}
	\includegraphics[width=0.49\textwidth]{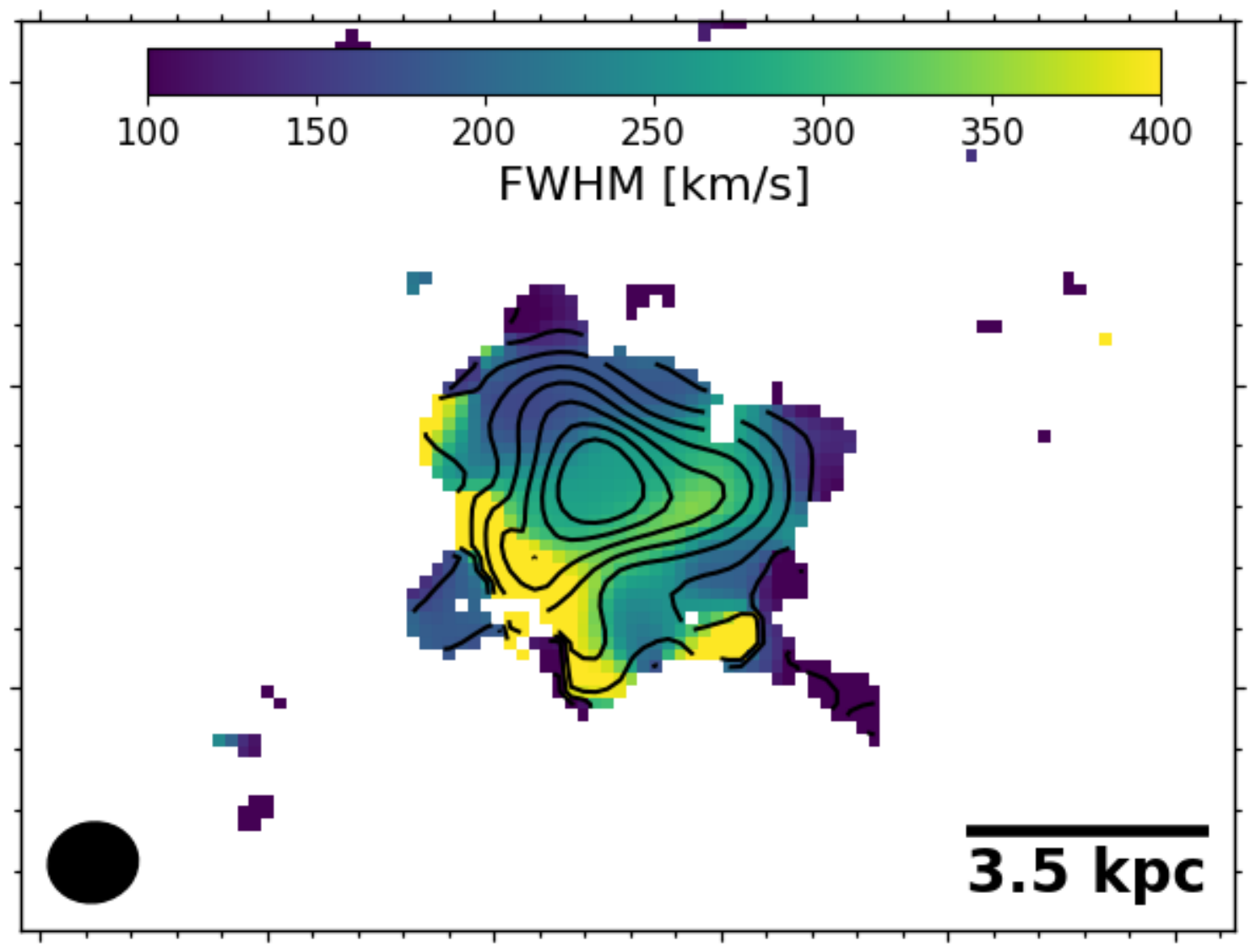}
	\caption{Maps of velocity centroid and line FWHM determined by the pixel-by-pixel fitting scheme discussed in Section \ref{sec:kinematics}. These images have the same field-of-view as the CO(1-0) flux map in Fig. \ref{fig:comparison}. The solid black ellipse is the synthesized beam.
	}
	\label{fig:maps}
\end{figure}

Maps of integrated flux, velocity, and FWHM of the CO(1-0) line were created by fitting the spectra extracted from individual pixels within the data cube. Each spectrum was averaged over a box the size of the synthesized beam. We allowed for the possibility of multiple coincident velocity structures by fitting each spectrum with up to two Gaussian components. The significance of each velocity component was tested using a Monte Carlo analysis employing 10000 iterations, with a detection requiring $5\sigma$ significance. The presence of one component was required before attempting to fit a second. We found that a single Gaussian component was sufficient to accurately model the spectrum for each pixel. Instrumental broadening has been incorporated into the model.

The integrated flux map is presented in the lower left panel of Fig. \ref{fig:comparison}, and the velocity centroid and full-width at half maximum (FWHM) maps are presented in Fig. \ref{fig:maps}. The molecular gas distribution is compact, with all of the detected gas confined to a $1''$ ($4.35\kpc$) radius. It is marginally resolved into three extensions to the NE, SE, and W. These extensions do not exhibit any clear velocity gradients. The emission is at its most blueshifted ($-150\kmps$) at the central peak and in the farthest extent of the SE extension. The NE and W extensions have more moderate velocities ($0$ to $50\kmps$), but the velocity transition to the central value occurs over the span of about a beam, so it is likely the result of the resolution element smearing two distinct velocities together. The linewidths, which range from $100$ to $400\kmps$ FWHM, are broader than those seen along filaments in other systems ($<100\kmps$) \citep[e.g.][]{Russell16, Vantyghem16, Vantyghem18}, consistent with a superposition of gas structures at a range of velocities.

{\it Chandra} X-ray, {\it HST} WFPC2 F606W, Ly-$\alpha$ ({\it HST} ACS/SBC F140LP), and FUV ({\it HST} ACS/SBC F165LP) images \citep{ODea10, Tremblay15} are shown alongside the CO(1-0) total flux map in Fig. \ref{fig:comparison}. The box in the X-ray image corresponds to the fields-of-view for the {\it HST} images, while the ALMA field-of-view is shown in the {\it HST} images. 
The central galaxy contains bright knots of nebular emission and FUV flux from recent star formation, one of which is coincident with the molecular gas. A brighter Ly-$\alpha$ and FUV peak is located $3.25\kpc$ ESE of the molecular gas peak. No molecular gas is present at this position; the CO emission truncates where the Ly-$\alpha$ and FUV fluxes begin to increase. Dust attenuation may account for this anti-correlation between the molecular gas and FUV emission.
Features that may be indicative of a merger remnant are seen in the {\it HST} F606W image. The most prominent clumps and filament are identified in Fig. \ref{fig:comparison}. However, this filter also includes H$\beta$ and [O\textsc{iii}] line emission, which may account for some of the observed features.

The molecular gas is offset by $10\kpc$ to the south of the peak in X-ray emission, which itself does not coincide with structure at any other wavelength. The bright X-ray emission in the cluster core arcs around the position of the CO(1-0) peak. In addition, the X-ray atmosphere contains a series of depressions that likely correspond to cavities inflated by the central AGN (see Section \ref{sec:cavities}).

\begin{figure}
	\includegraphics[width=\columnwidth]{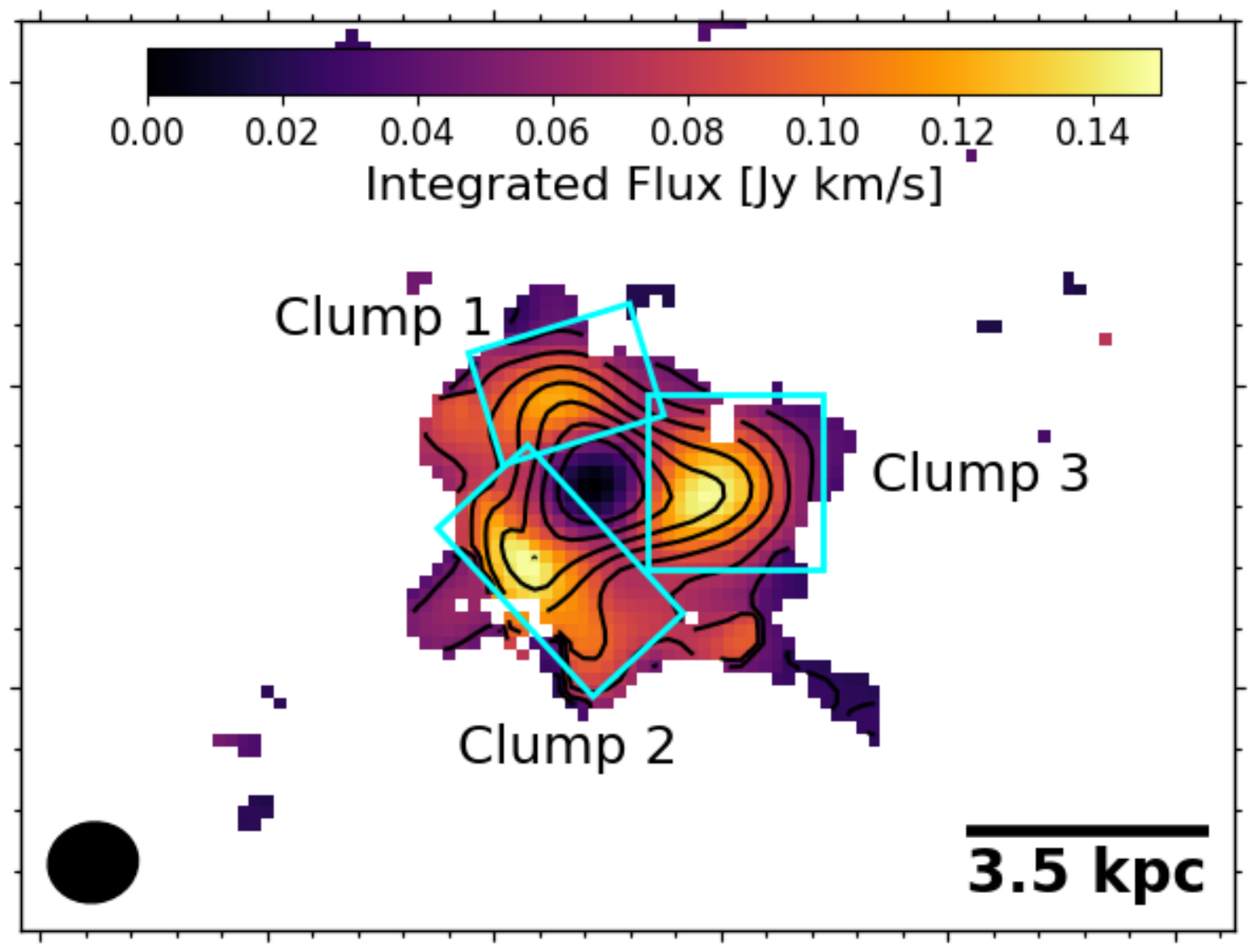}
	\caption{Maps of the integrated CO(1-0) flux after subtracting a point source centered on the peak of the gas distribution. The contours correspond to the original distribution of integrated flux, as shown in Fig. \ref{fig:maps}.}
	\label{fig:beamsub}
\end{figure}

The molecular extensions can be better visualized by subtracting the contribution from an unresolved point source coincident with the CO(1-0) peak. This is shown in Fig. \ref{fig:beamsub}. First we created an image of the beam centered on the pixel with the largest integrated flux. This image was further convolved with a beam-shaped tophat kernel, as the pixel-based spectral fitting used to create the maps extracts the emission from a region of this shape. The image was then scaled to match the flux of the brightest pixel and subtracted from the integrated flux map. The unresolved emission subtracted from the integrated flux map amounted to 22\% of the total flux.

\begin{figure}
	\includegraphics[width=\columnwidth]{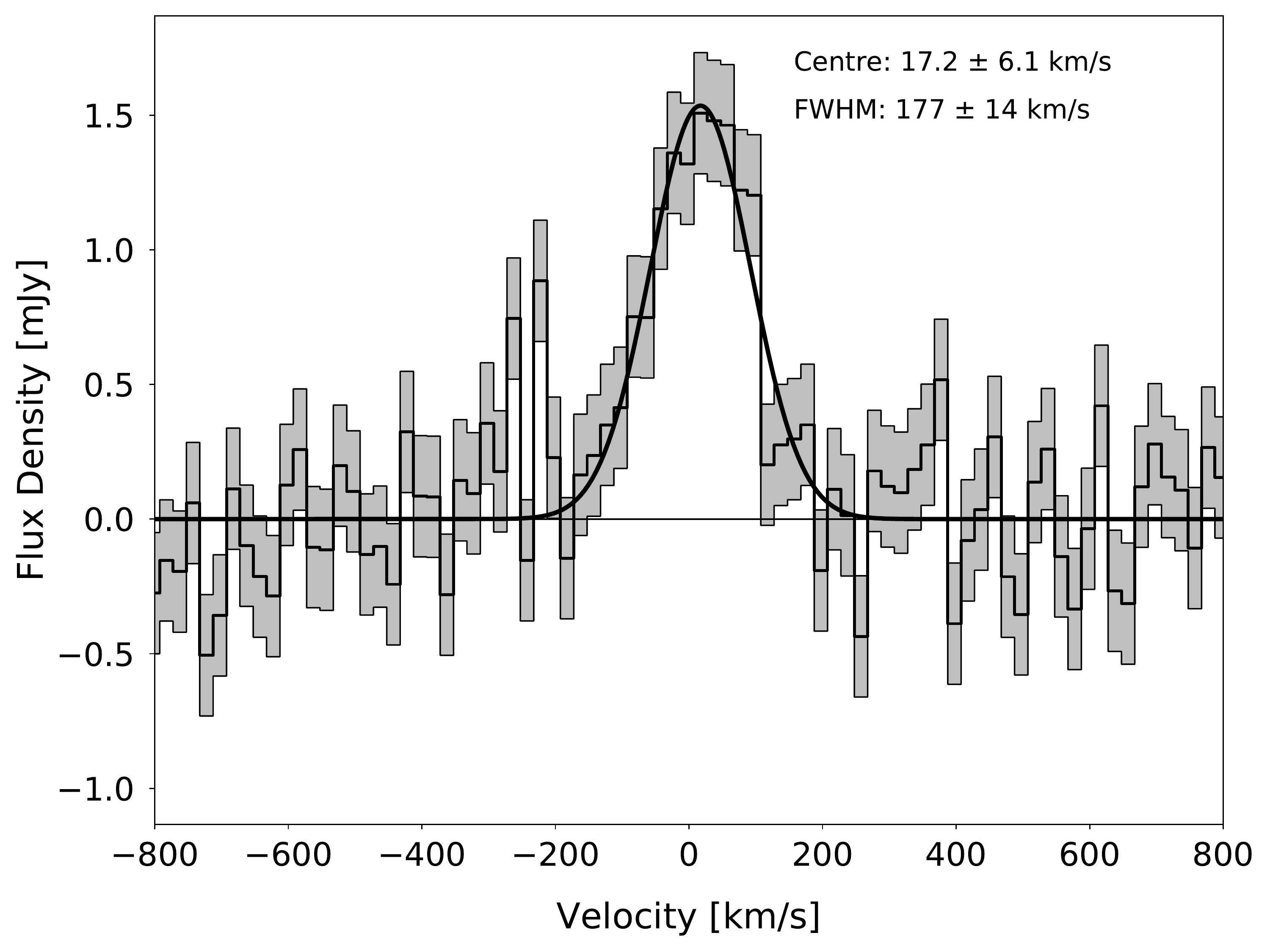}
	\includegraphics[width=\columnwidth]{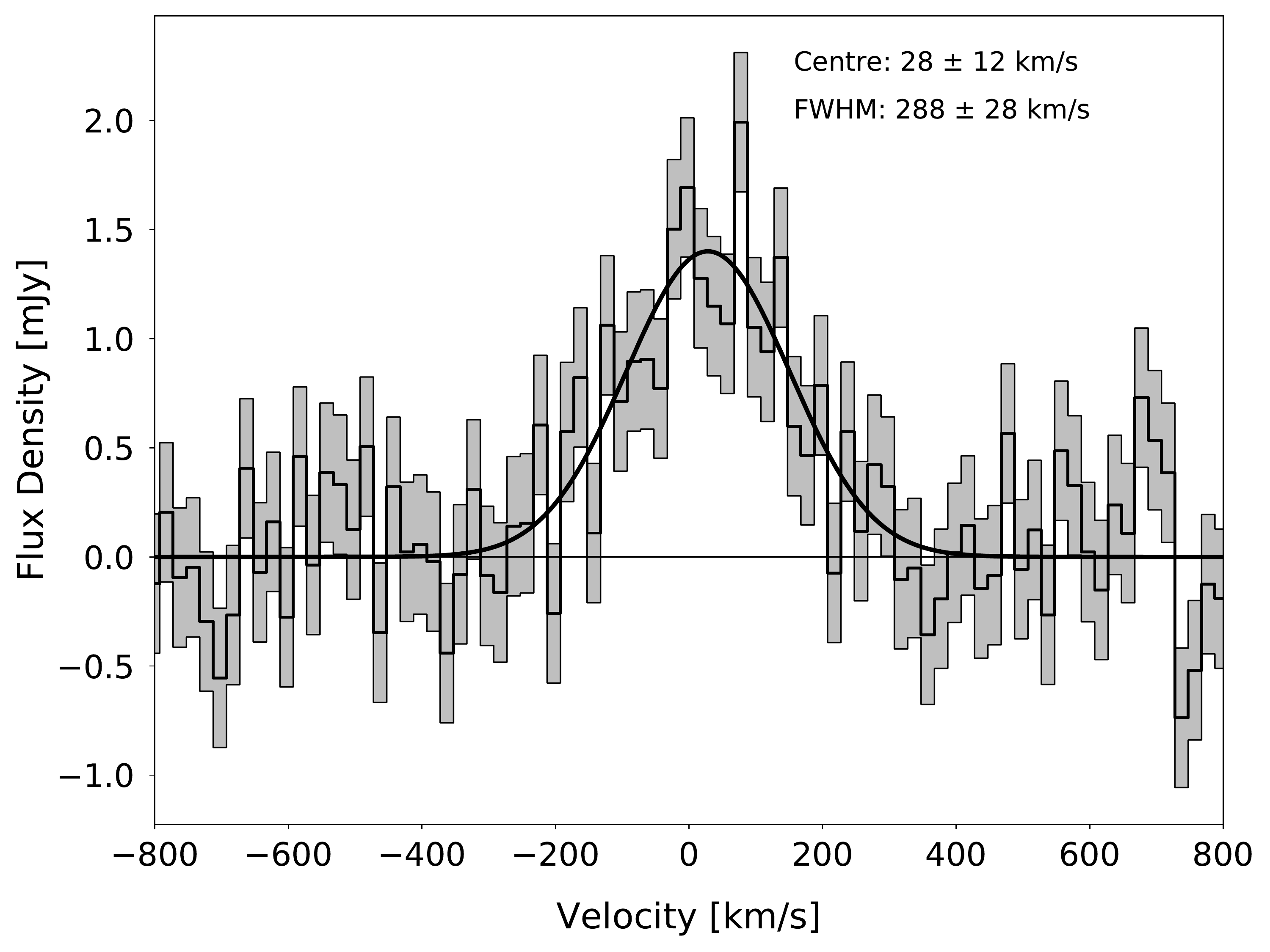}
	\includegraphics[width=\columnwidth]{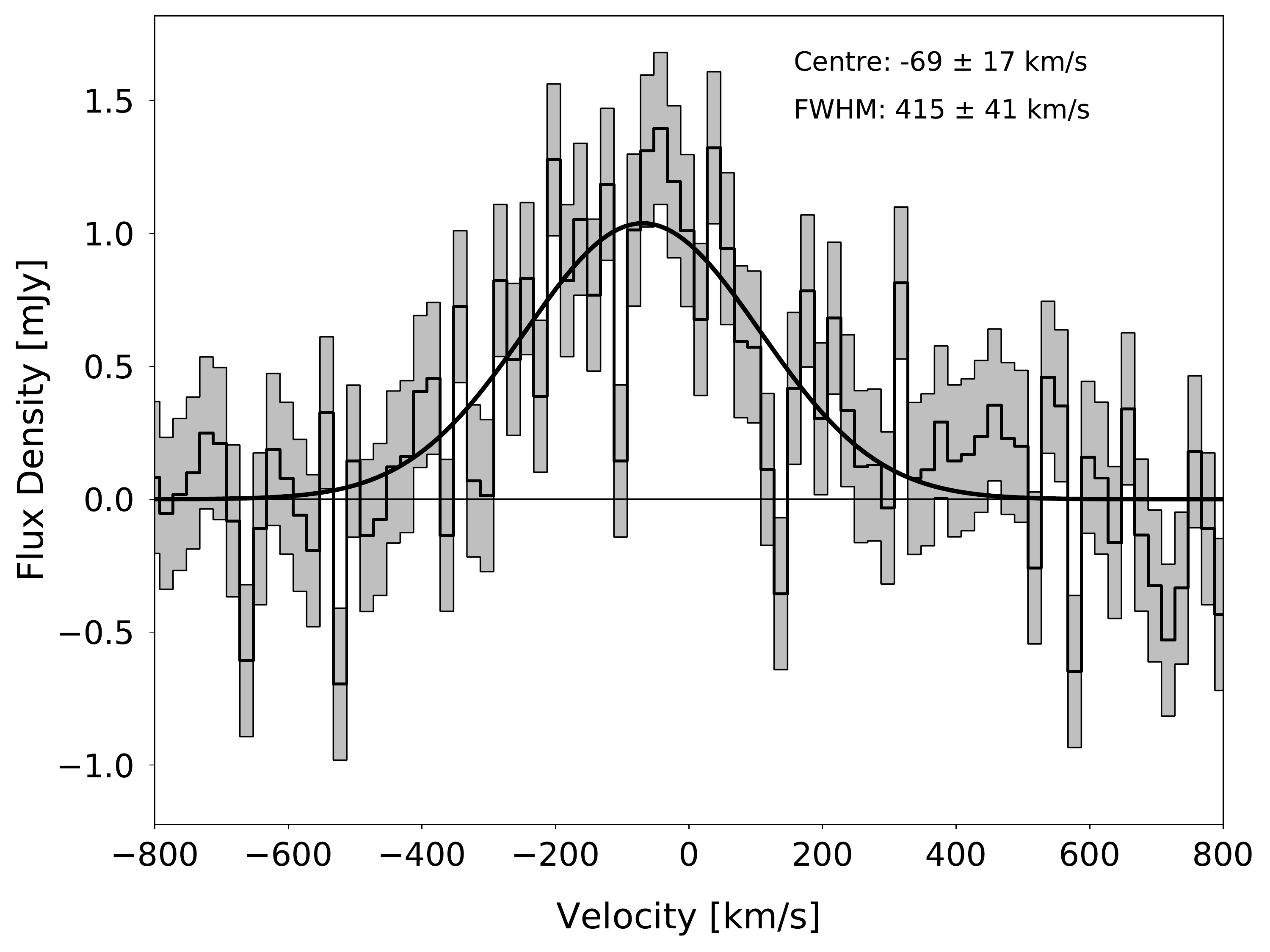}
	\caption{CO(1-0) spectral to the three clumps identified in Fig. \ref{fig:beamsub}.}
	\label{fig:clump_spectra}
\end{figure}

The resulting map shows that the gas morphology is separated into the three extensions, although none of them are resolved. They have been identified in Fig. \ref{fig:beamsub}. The corresponding spectra are shown in Fig. \ref{fig:clump_spectra}, and their spectral fits are listed in Table \ref{tab:spectra}.

\begin{figure}
	\includegraphics[width=\columnwidth]{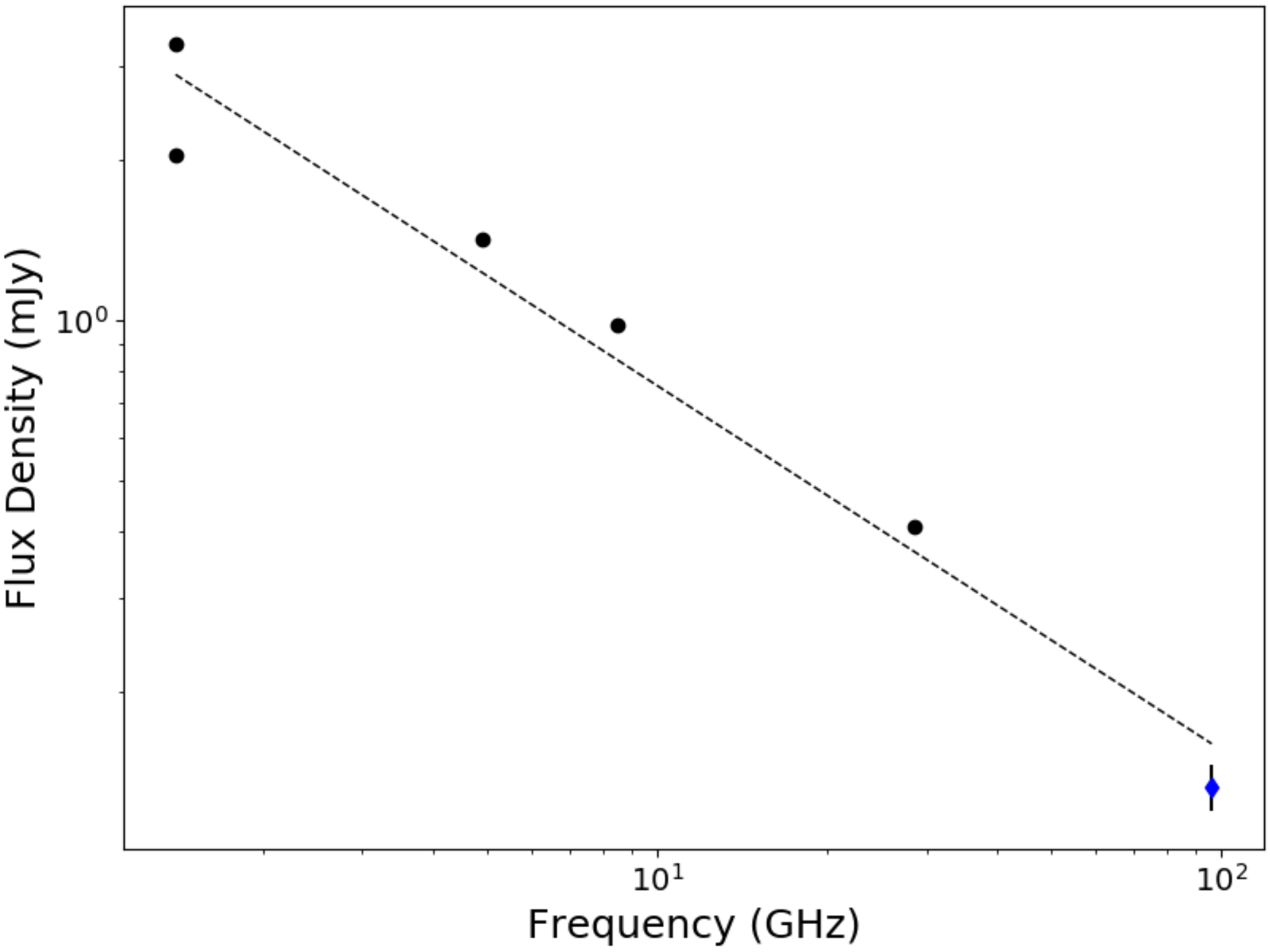}
	\caption{Radio spectral energy distribution, with the ALMA continuum source measurement shown in blue. The best-fitting power law, indicated by the dashed line, has a slope of $-0.685\pm0.063$.}
	\label{fig:sed}
\end{figure}

\subsection{Millimetre Continuum}
\label{sec:continuum}

An image of the continuum at $96.193\GHz$ ($3.1166\mm$) was created by imaging the line-free channels in a square region measuring $40''$ on each side. A single, unresolved source was identified (R.A.: 10:23:39.5948, decl.: +4:11:11.032), which had a total integrated flux density of $133\pm13 \uJy$. The contours for this source are shown in red Fig. \ref{fig:comparison}. These contours begin at $3\sigma$, where $\sigma=7.45\uJypbeam$, and increase in $4\sigma$ steps. A two-dimensional gaussian fit to the continuum flux yielded an accuracy for the centroid of $0.006$ arcseconds. We searched for, but did not detect, evidence for absorption against the continuum source using the native velocity resolution of the observations.

In Fig. \ref{fig:sed} we plot the ALMA continuum alongside flux measurements at lower frequencies, which were obtained from either the VLA or BIMA \citep{Egami06, Coble07, Giacintucci14}. Error bars are included, but are smaller than the markers for all measurements except those provided in this work. The spectral energy distribution (SED) is best fit by a power law with spectral index\footnote{Following the convention $S_{\nu} \propto \nu^{-\alpha}$} $\alpha = 0.685\pm0.063$, which is a typical value for an AGN \citep[e.g.][]{Hogan15a}.

\begin{figure}
	\includegraphics[width=\columnwidth]{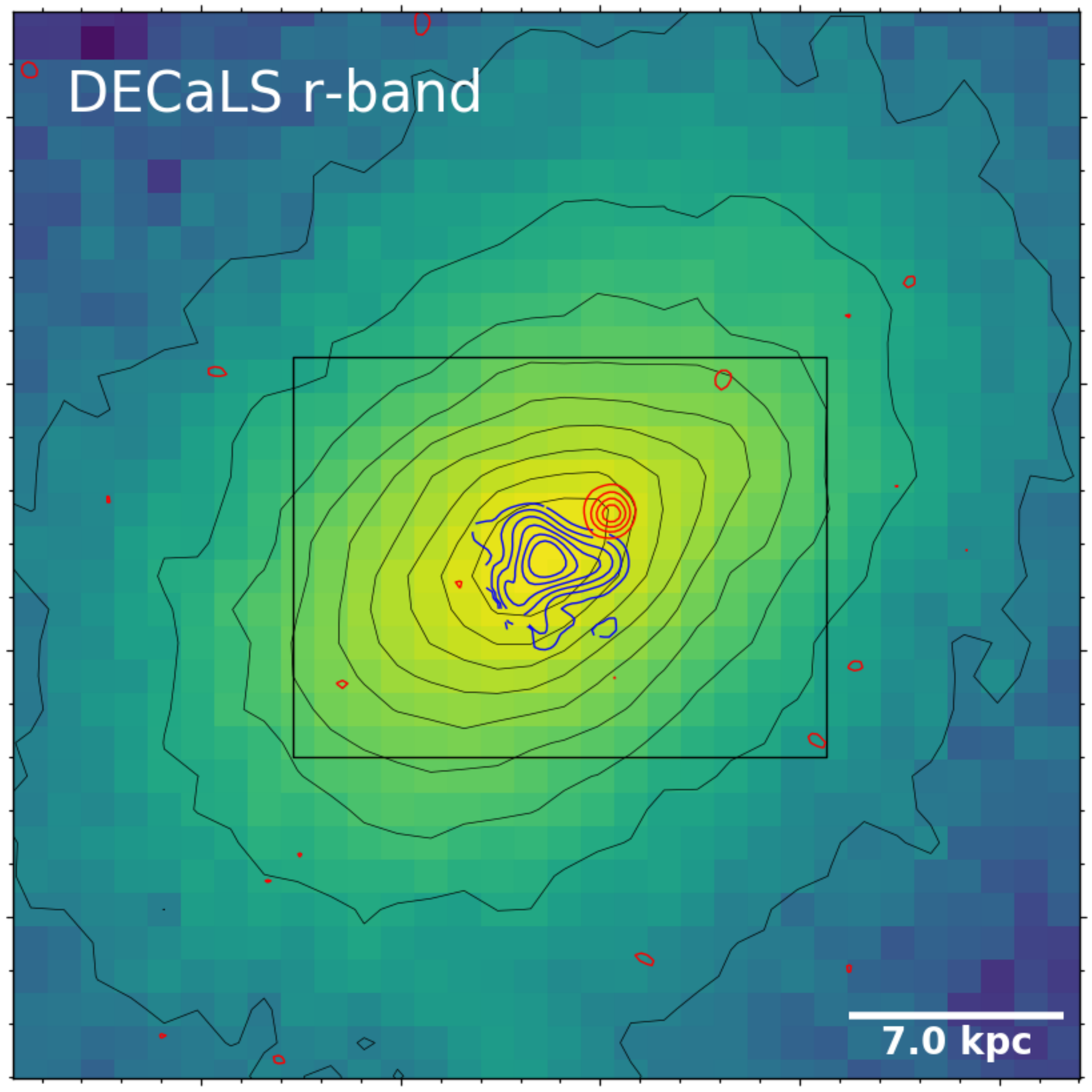}
	\caption{DECaLS r-band image of Zw~3146 in the same field-of-view as the HST F606W image in Fig. \ref{fig:comparison}. Contours of the ALMA CO(1-0) emission and continuum source are shown in blue and red, respectively. Te black contours correspond to the DECaLS r-band flux. The black rectangle indicates the ALMA CO(1-0) field-of-view as shown in Fig. \ref{fig:comparison}.}
	\label{fig:decals}
\end{figure}

\subsection{The Molecular Gas-AGN Offset}
\label{sec:offset}

As is evident in Fig. \ref{fig:comparison}, the molecular gas and continuum source are mutually offset. The two peaks are $2.6\kpc$ ($0.6''$) apart, and only a minimal overlap is present in the fainter reaches of the two distributions.
This offset cannot originate from a phase calibration issue, as the continuum and line measurements are obtained from the same observation. 

Independent observations of the radio continuum at lower resolution confirm the position of the $\mm$ continuum source identified here. In a VLA 4.9 GHz observation the continuum centroid is $0.3''$ SSW of the ALMA continuum source \citep{ODea10}. Though the coarse resolution ($1.4''\times 0.9''$) prevents the CO(1-0) centroid and continuum source from being definitively distinguished, the VLA centroid is closer to the ALMA continuum position. Note that the VLA coordinates were reported in the B1950 equinox, so have been converted to J2000. \citet{Giacintucci14} also report a continuum centroid corresponding to VLA 4.9 and 8.5 GH observations in the C configuration. The position is consistent with the location of the ALMA continuum source, although no uncertainties are provided and the beam sizes are larger. Nevertheless, these independent radio observations indicate that the ALMA continuum position reported here is likely accurate.

The stellar continuum is difficult to disentangle from nebular emission in the HST images. In order to determine the position of the galactic center, we obtained a Dark Energy Camera Legacy Survey ({\it DECaLS}) $r$-band image using the legacy skyviewer\footnote{https://www.legacysurvey.org/viewer}, which is shown in Fig. \ref{fig:decals}. This image is uncontaminated by nebular line emission and stellar features from a merger remnant, providing a smooth tracer of the stellar continuum. Contours of molecular line emission and continuum source are overlaid in blue and red, respectively, in Fig \ref{fig:decals}, along with black contours tracing the stellar emission. It is the molecular gas that is coincident with the galactic center, not the continuum emission from the AGN.

All taken together, these results imply that the molecular gas resides at the galactic center while the continuum source is offset by $2.6\kpc$. The molecular gas is coincident with both the stellar centroid and bright knots of nebular emission and star formation, while the continuum position is consistent with independent measurements at lower frequencies. This correspondence spanning multiple types of observations indicates that the astrometry of the ALMA observation is reliable. We explore possible physical interpretations for this offset in Section \ref{sec:offsetorigin}.

\subsection{X-ray Cavities}
\label{sec:cavities}

\begin{figure}
	\includegraphics[width=\columnwidth]{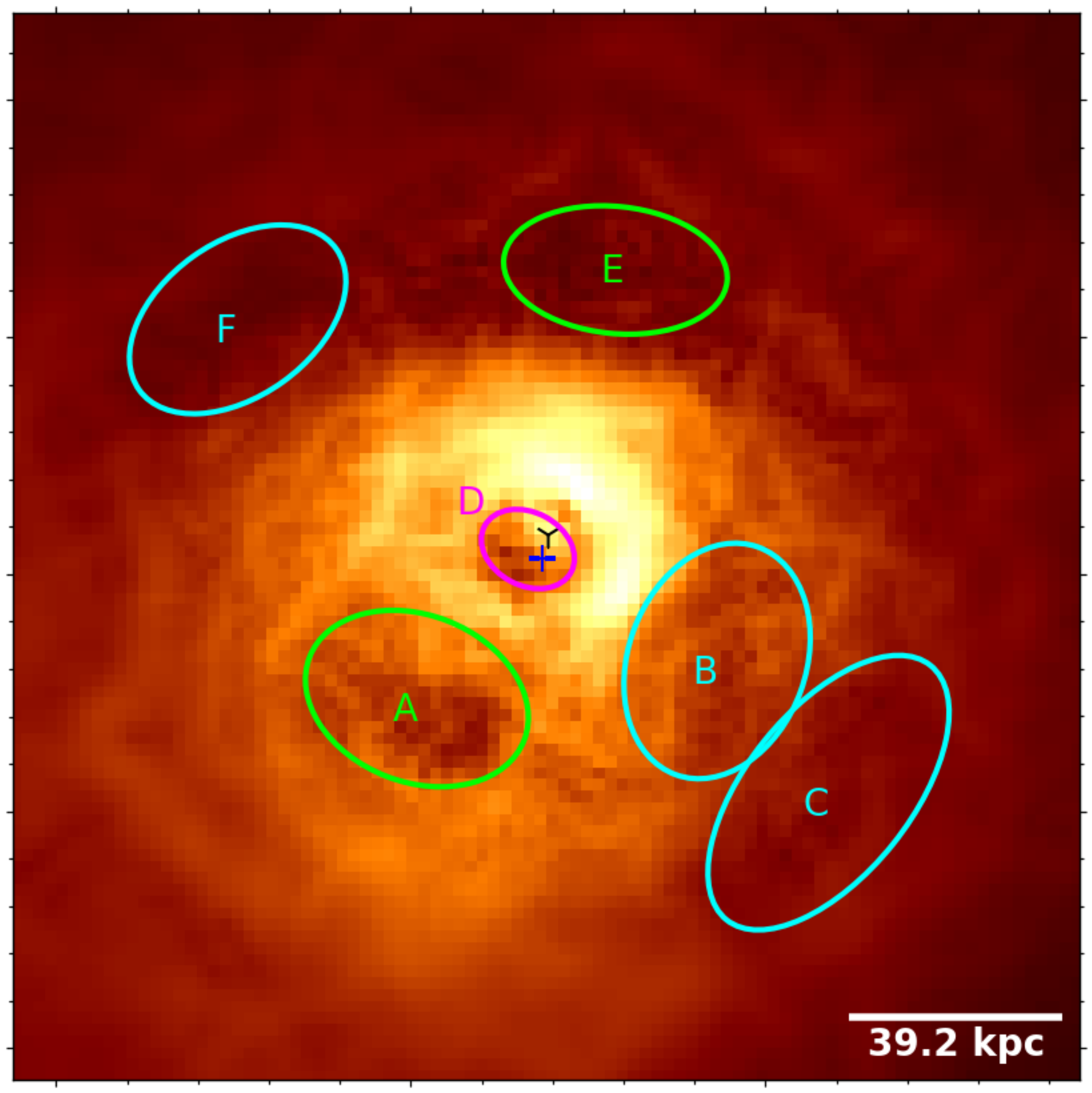}
	\caption{X-ray image processed using the GGM algorithm to enhance surface brightness edges, shown with the same field-of-view as the X-ray image in Fig. \ref{fig:comparison}. Six candidate X-ray cavities have been identified. The location of the cluster center is indicated by the black "Y". The blue $+$ indicates the molecular gas centroid.}
	\label{fig:ggm}
\end{figure}

In order to investigate any possible connection between the molecular gas and X-ray cavities, we present a new search for cavities in Zw~3146 using the {\it Chandra} data reduced by \citet{Hogan17a} and \citet{Pulido18}. The $0.5-7\keV$ X-ray surface brightness image is shown in Fig. \ref{fig:comparison}. In addition, we have processed this image using the gaussian gradient magnitude algorithm \citep[GGM;][]{Sanders16a, Sanders16b} with a radially-dependent filtering kernel. This algorithm enhances edges in surface brightness. The resulting image is shown in Fig. \ref{fig:ggm}, sharing a field-of-view with the X-ray image in Fig. \ref{fig:comparison}. The X-ray atmosphere contains a series of surface brightness depressions and edges corresponding to X-ray cavities and cold fronts, respectively \citep{Forman02, Rafferty06}.

We manually identified six candidate X-ray cavities using a combination of the GGM filtered image and original X-ray image. The sizes were determined qualitatively by estimating the size of the surface brightness depressions, using the edges identified in the GGM image as a guide. The cavities are shown and labelled in Fig. \ref{fig:ggm}. Two cavities were previously reported by \citet{Rafferty06}, though their work used a shallower X-ray image. Judging from the radial distances and sizes of these cavities, one corresponds directly to Cavity "A". The other cavity reported by Rafferty et al. is identified as two in our work -- "B" and "C".
Cavity "D" is located close to the cluster center and may correspond to motions unrelated to AGN activity, but we tabulate its derived properties nonetheless.
Cavities "E" and "F" are located outside of a cold front and are less well-defined than their counterparts.

A list of all cavity parameters can be found in Table \ref{tab:cavities}. 
A qualitative figure of merit (FOM), which describes a cavity's contrast with its surroundings, was provided for each cavity \citep{Rafferty06}. High contrast cavities (${\rm FOM}=1$) are surrounded by bright rims, medium contrast (${\rm FOM}=2$) are partially surrounded by bright rims, and low contrast (${\rm FOM}=3$) have no or faint rims.
All necessary thermodynamic quantities were obtained by interpolating the deprojected radial profiles from \citet{Pulido18}. Note that the proximity of cavity D to the cluster center has a strong systematic effect on the cavity age, and therefore its power. The cluster centroid was determined from an aperture with a $20''$ ($87\kpc$) radius. It is offset from the X-ray peak by $6\kpc$.

The enthalpy required to inflate a cavity filled with relativistic gas is given by $E_{\rm cav}=4pV$, where $p$ is the cavity pressure and $V$ is its volume. The cavity was assumed to be in pressure balance with the surrounding gas. The confining pressure was interpolated from the deprojected pressure profile to a radius equal to the distance to the cavity center ($R$).

The cavity volume was estimated as the geometric mean of the volumes of oblate and prolate geometries, with semiaxes $a$ and $b$ matching those of the fitted ellipse, so that $V=\frac{4}{3}\pi(ab)^{3/2}$. This is equivalent to assuming that the length of the semiaxis along the line-of-sight is $r=\sqrt{ab}$. The uncertainty for each cavity volume was determined in two ways. First, a $15\%$ uncertainty was assumed in both $a$ and $b$ and propagated. 
Second, we used the prolate and oblate geometries as measures of the minimum and maximum volumes, respectively. The volume uncertainty was then taken to be the difference between maximum (minimum) volume and cavity volume, in absolute value. This evaluates to a fractional uncertainty of $\sqrt{a/b}-1$ for an oblate geometry and $1-\sqrt{b/a}$ for prolate, which are respectively the positive and negative errors. This is the dominant uncertainty when $a/b \geq 1.74$ when compared to a $15\%$ uncertainty in both $a$ and $b$.

The cavity age was estimated using both the sound crossing time and buoyant rise time \citep{Birzan04}. The sound crossing time, which assumes the cavity rises at the speed of sound ($c_s$), is simply a function of ICM temperature, and is given by
\begin{equation}
t_{c_s} = R/c_s = R\sqrt{\mu m_H / \gamma kT},
\end{equation} 
where $\gamma=5/3$ is the ratio of specific heats for an ideal gas, and $\mu=0.6$ is the mean particle weight for the ICM.
The buoyant rise time is the time taken for the cavity to rise to its current projected distance at its terminal velocity ($v_t$),
\begin{equation}
t_{\rm buoy} = R/v_t \simeq R\sqrt{SC/2gV}.
\end{equation}
Here $C=0.75$ is the drag coefficient \citep{Churazov01} and $S$ is the bubble cross-section, which is assumed to be its projected area. 
The acceleration under gravity at radius $R$ was determined using the cluster mass profile from \citet{Pulido18}. 
The two timescales agree to within $20\%$ for all cavities. The buoyancy time was used to determine the mean cavity power, $P_{\rm cav} = E_{\rm cav}/t_{\rm buoy}$.

These quantities were then used to place mass constraints on the AGN outbursts. The first, $M_{\rm acc} = E_{\rm cav}/\epsilon c^2$, is the mass that must be accreted onto the black hole to fuel the outburst at an efficiency of $\epsilon$, which we assume to be $0.1$. The total accreted gas mass of $4.6\e{7}\Msun$ can easily be supplied by the observed $5\e{10}\Msun$ of molecular gas. The corresponding accretion rate, $\dot{M}_{\rm acc}=P_{\rm cav}/\epsilon c^2$, ranges from $\sim 0.1-0.3\Msunpyr$ for each cavity. 
The other interesting mass is that of hot gas displaced by the cavities, as this mass dictates the maximum mass that can be lifted by the cavities. The mass of displaced gas is given by $M_{\rm disp} = n\mu m_H V$, where $n=n_e+n_H$ is the density of the surrounding gas. Each of the three larger cavities can lift a few $10^{10}\Msun$, so the observed molecular gas supply could have been lifted by the cavities.

\begin{table*}
\caption{Cavity Measurements}
\begin{center}
\begin{tabular}{l c c c c c c c c c c c}
\hline \hline
Cavity  &  FOM$^a$ & Major  &  Minor  &  $R$  &  $4pV$  &  $t_{cs}$  &  $t_{\rm buoy}$  &  $P_{\rm cav}$  &  $M_{\rm acc}$  &  $\dot{M}_{\rm acc}$  & $M_{\rm disp}$  \\
  &  &  $\kpc$  &  $\kpc$  &  $\kpc$  &  $10^{58}\erg$  &  $\Myr$  &  $\Myr$  &  $10^{44}\ergps$  &  $10^{6}\Msun$  &  $\Msunpyr$  &  $10^{10}\Msun$  \\
\hline
A & 1 & $21.1$  &  $15.3$  &  $41.7$  &  $161^{+58}_{-56}$  &  $38^{+3.2}_{-2.5}$  &  $41.2\pm2.8$  &  $12.4^{+4.6}_{-4.4}$  &  $9.02^{+3.3}_{-3.1}$  &  $0.219^{+0.081}_{-0.077}$  &  $2.69\pm0.85$ \\
B & 2 & $22.2$  &  $16.3$  &  $38.5$  &  $214^{+77}_{-74}$  &  $34^{+2.8}_{-2.2}$  &  $36.3\pm2.4$  &  $18.7^{+6.9}_{-6.6}$  &  $12^{+4.3}_{-4.2}$  &  $0.33\pm0.12$  &  $3.35\pm1.1$ \\
C & 2 & $30$  &  $15.2$  &  $70$  &  $204^{+86}_{-68}$  &  $57.5^{+3.2}_{-2.6}$  &  $71.2^{+4.4}_{-4.3}$  &  $9.08^{+3.9}_{-3.1}$  &  $11.4^{+4.8}_{-3.8}$  &  $0.16^{+0.068}_{-0.054}$  &  $2.76^{+1.1}_{-0.88}$ \\
D & 1 & $9.01$  &  $6.72$  &  $7.65$  &  $30.6\pm11$  &  $8.81^{+0.7}_{-0.59}$  &  $6.68^{+0.62}_{-0.6}$  &  $14.5^{+5.5}_{-5.3}$  &  $1.71^{+0.63}_{-0.61}$  &  $0.257^{+0.097}_{-0.094}$  &  $0.815\pm0.26$ \\
E & 3 & $20.6$  &  $11.7$  &  $47.5$  &  $86.6^{+31}_{-29}$  &  $45.2^{+2.9}_{-2.3}$  &  $52.1\pm3.3$  &  $5.27^{+1.9}_{-1.8}$  &  $4.85^{+1.7}_{-1.6}$  &  $0.0931^{+0.034}_{-0.032}$  &  $1.58^{+0.52}_{-0.5}$ \\
F & 3 & $22.3$  &  $14.2$  &  $70$  &  $118^{+40}_{-39}$  &  $57.5^{+3.2}_{-2.6}$  &  $77.9^{+4.8}_{-4.7}$  &  $4.8\pm1.6$  &  $6.61\pm2.2$  &  $0.0847\pm0.029$  &  $1.6\pm0.51$ \\
\hline \hline
\end{tabular}
\end{center}
Notes: $^{a}$ The cavity figure of merit, which is a qualitative measure of the contrast between a cavity and its surroundings: (1) high contrast, bright rim surrounds cavity; (2) medium contrast, bright rim partially surrounds cavity; and (3) low contrast, no rim, or faint rim surrounds cavity.
\label{tab:cavities}
\end{table*}

\section{Discussion}

\subsection{Origin of the Molecular Gas-AGN Offset}
\label{sec:offsetorigin}

As discussed in Section \ref{sec:offset}, the molecular gas and continuum source are mutually offset. The molecular gas resides at the galactic center, while the continuum source is located $2.6\kpc$ to the NW. In this Section we discuss potential causes for this offset. 
The first two attribute the continuum emission to a source other than the SMBH of the BCG, namely emission from a single lobe of a radio jet or the nucleus of a secondary galaxy. The latter two -- gravitational wave recoil and acceleration by an asymmetric radio jet -- are mechanisms that would naturally kick or accelerate the black hole from the galactic center.

Offsets between SMBHs and their galactic centers have been observed in other systems \citep[see][for an extensive list]{Bartlett20}. Most common are the $\sim10-100\pc$ displacements, which are consistent with an oscillating SMBH following a gravitational wave recoil \citep[e.g.][]{Lena14}. M87 is an example of such a system, as its SMBH is offset from the photo-center by $6.8\pc$ \citep{Batcheldor10}. More exceptional displacements are also observed. The radio-loud quasar 3C~186, for example, harbors an $11\kpc$ displacement with a corresponding velocity of $-2170\kmps$ \citep{Chiaberge17}.

\subsubsection{Emission from a lobe}

Instead of originating from the supermassive black hole's accretion disk, the detected continuum emission could instead originate from the lobe of a radio jet. This would naturally explain the offset, although it requires that the emission from both the core and a second lobe are undetected. Given the RMS in the continuum image of $7.5\uJypbeam$, any source brighter than $22.5\uJy$ would be detected in our observations. 

The continuum emission, with a spectral index of $0.7$, is consistent with originating from an AGN. Spectral indices steeper than $0.5$ are generally associated with non-cores, which include lobes, mini-halos, and relics \citep{Hogan15a}. If the continuum emission arises from a radio lobe, then it is poorly associated with the X-ray cavities. The continuum emission that would correspond to the putative lobe is located NW of the galactic center, while the most prominent cavities are either to the SE or SW. Cavity D is coincident with the continuum emission, although its overlap with the galactic center makes it difficult to interpret the direction in which the jet would have been launched.

\subsubsection{The SMBH of a secondary galaxy}
\label{sec:secondary_galaxy}

The spatially offset continuum emission could also originate from the SMBH in a secondary galaxy that is projected near the BCG core. This SMBH could either be in an independent background galaxy or the center of a galaxy merging with the BCG.

The probability that the continuum source is a chance alignment with an unrelated background radio source is low, based on our estimates of the source density of radio continuum sources at 100 GHz. Lacking a $100\GHz$ survey sensitive enough to detect the source density down to $0.13\mJy$, we have instead used a $31\GHz$ survey of known extragalactic sources \citep{Mason09}. A source density of $16.7\deg^{-2}$ is appropriate for sources with flux densities $\geq 1\mJy$ at $31\GHz$, which translates to a flux of $0.6\mJy$ at $100\GHz$, assuming a spectral index of 0.4 \citep{Marriage11}. We then randomly populated 1000 fields each with a source density of $16.7\deg^{-2}$. This exercise results in a probability of $1.2\e{-6}$ that two sources are separated by $<0.6''$. While our assumed density is based on a flux limit of $1\mJy$, which is higher than the Zw~3146 continuum source flux ($0.133\mJy$), the actual source density would need to be several orders of magnitude larger to exceed a probability of close but random coincidence of even a few percent.

The likelihood that the continuum source originates from the core of a merger remnant is much higher. Clumps and trails possibly corresponding to stellar emission from a merger remnant are detected in the {\it HST} F606W image (Fig. \ref{fig:comparison}). H$\beta$ and [O{\sc iii}] line emission may also contribute to these features. Unlike in systems with dual nuclei, we do not detect excess optical emission at the position of the continuum source.
A similar conclusion was drawn in Zw~8193, where the unresolved radio emission is offset from the center of the BCG by 3 arcsec and may be associated with a merging galaxy \citep{ODea10}.

\subsubsection{Gravitational wave recoil}

As two galaxies merge, their black holes migrate toward the center of the gravitational potential through dynamical friction \citep{Begelman80}. 
The resulting binary system tightens through interactions with nearby stars or gravitational drag from gas until the SMBHs eventually coalesce \citep{Merritt05, Mayer07}. During the final stage of coalescence, the anisotropic emission of gravitational waves imparts a recoil velocity to the coalesced object \citep{Gonzalez07, Campanelli07}. This ``kick'' can in rare circumstances be as large as $4000\kmps$, so can result in large displacements between the resulting SMBH and the galactic center.

When a SMBH experiences a large recoil ($v_{\rm kick} \sim 40-90\%$ of the escape velocity) in an elliptical galaxy potential, its subsequent motion is characterized by three phases \citep{Gualandris08}. First, the large-scale perturbations -- those exceeding the core radius ($r_c \sim 1\kpc$) -- are damped within $\sim 10^7\yr$. Then, when the amplitude becomes comparable to the galaxy's core radius, the oscillations of the SMBH persist for $\sim 1\Gyr$. Finally, the SMBH reaches thermal equilibrium with the stars, at which point the oscillations are negligibly small.

The large ($2.6\kpc$) displacement observed in Zw~3146 implies that any SMBH coalescence would have occurred recently -- within the last $10^{7}\yr$. In a merger between two galaxies with $\sim 10^9\Msun$ SMBHs, coalescence occurs within $1-2\Gyr$ \citep{Khan12}. Stellar disturbances from a merger will persist for about $2\Gyr$ \citep{Lotz08}, but may be shorter lived in a rich cluster. 
The features observed in the {\it HST} F606W image (Fig. \ref{fig:comparison}) may correspond to tidal features induced by a merger. Thus, it is plausible that the SMBHs have recently coalesced and that gravitational wave recoil accounts for the displacement of the continuum source.

\subsubsection{Acceleration by an asymmetric radio jet}

If the radio jets are intrinsically asymmetric in their power output, then the SMBH will experience a net thrust that will displace it from its equilibrium position \citep{Shklovsky82}. 
When the black hole is accreting near the Eddington limit, the resulting thrust can cause it to oscillate within the host galaxy \citep{Wang92, Tsygan07, Kornreich08, Lena14}. Clusters are generally not expected to contain asymmetric jets, as even in the most powerful systems the SMBH resides at the center of the BCG and, on parsec scales, the jets are two-sided \citep{Liuzzo10}.
However, while most cavities in Zw~3146 are balanced by another on the opposite side of the BCG, the innermost cavity (cavity D) has no discernible counterpart. Since this cavity is positioned opposite the displacement direction, it can conceivably account for the displacement, provided it was inflated by an asymmetric jet.

The momentum flux for a one-sided relativistic jet is $P_{\rm jet}/c$. This is the reaction force applied to the black hole from which the jet originates. The resulting acceleration for a SMBH of mass $M_{\bullet}$ is
\begin{equation}
	a_{\bullet} = \frac{P_{\rm jet}}{M_{\bullet}c} \approx 1.7\e{-8} \left( \frac{P_{\rm jet}}{10^{45}\ergps} \right) \left( \frac{M_{\bullet}}{10^9\Msun} \right)^{-1} \cmpssq.
\end{equation}
The black hole mass in the Zw~3146 BCG, estimated using the 2MASS K-band magnitude, is $M_{\bullet} = 1.6\e{9} \Msun$ \citep{Main17}. 
The black hole acceleration for cavity D, using $P_{\rm cav}$ as an estimate for jet power (see Table \ref{tab:cavities}), evaluates to $1.5\e{-8}\cmpssq$.

A displaced black hole will experience a restoring acceleration from the gravitational pull of the BCG. Assuming the BCG is represented by a singular isothermal sphere with velocity dispersion $\sigma_{\ast}=300\kmps$, the gravitational acceleration ($2\sigma^2/R$) at a displacement of $R=1\kpc$ is $5.8\e{-7}\cmpssq$.
This is 40 times larger than the acceleration that would be imparted by cavity D assuming that it has no counter-jet. An asymmetric jet can therefore not account for the $2.6\kpc$ displacement of the continuum source.

\begin{figure*}
	\includegraphics[width=0.32\textwidth]{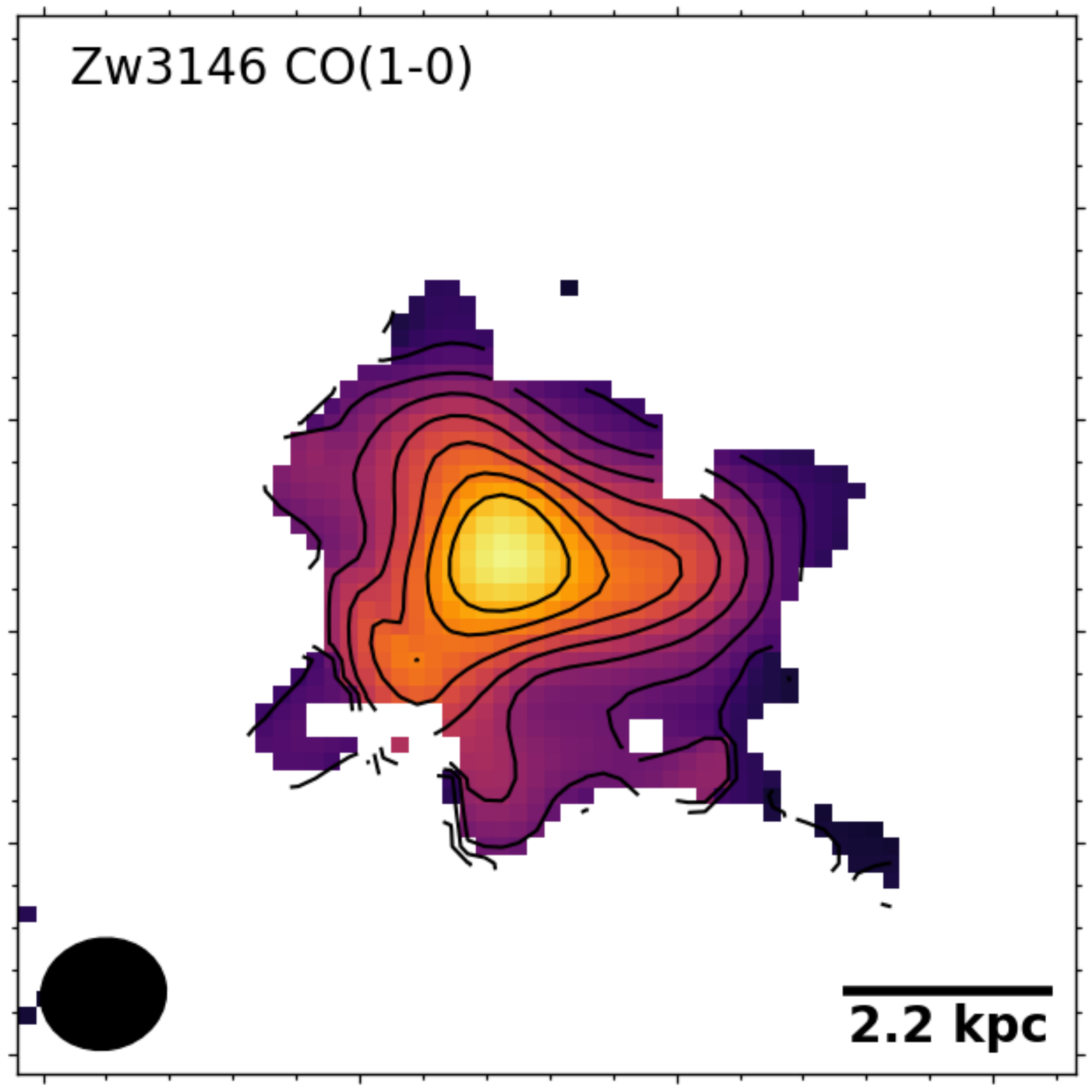}
	\includegraphics[width=0.32\textwidth]{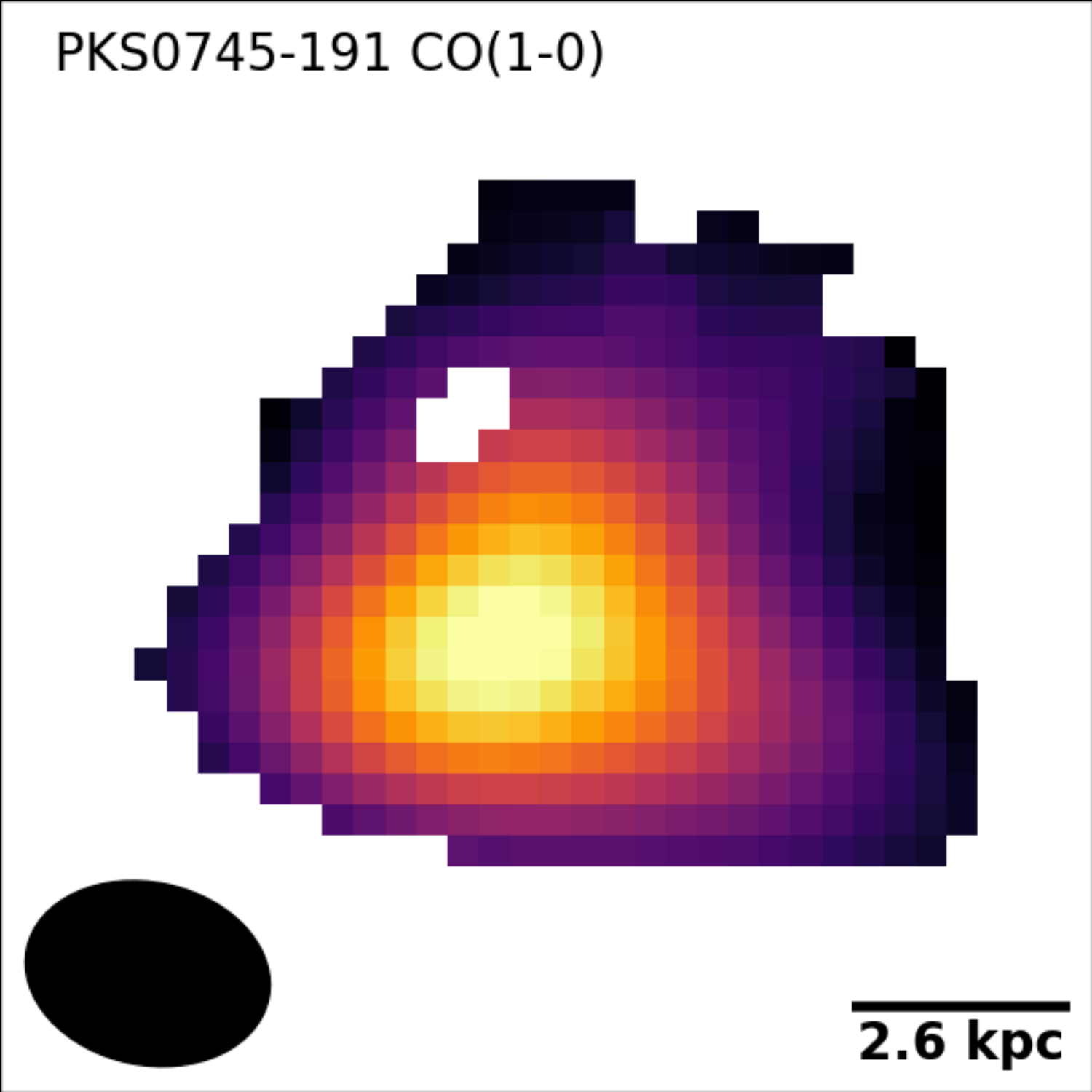}
	\includegraphics[width=0.32\textwidth]{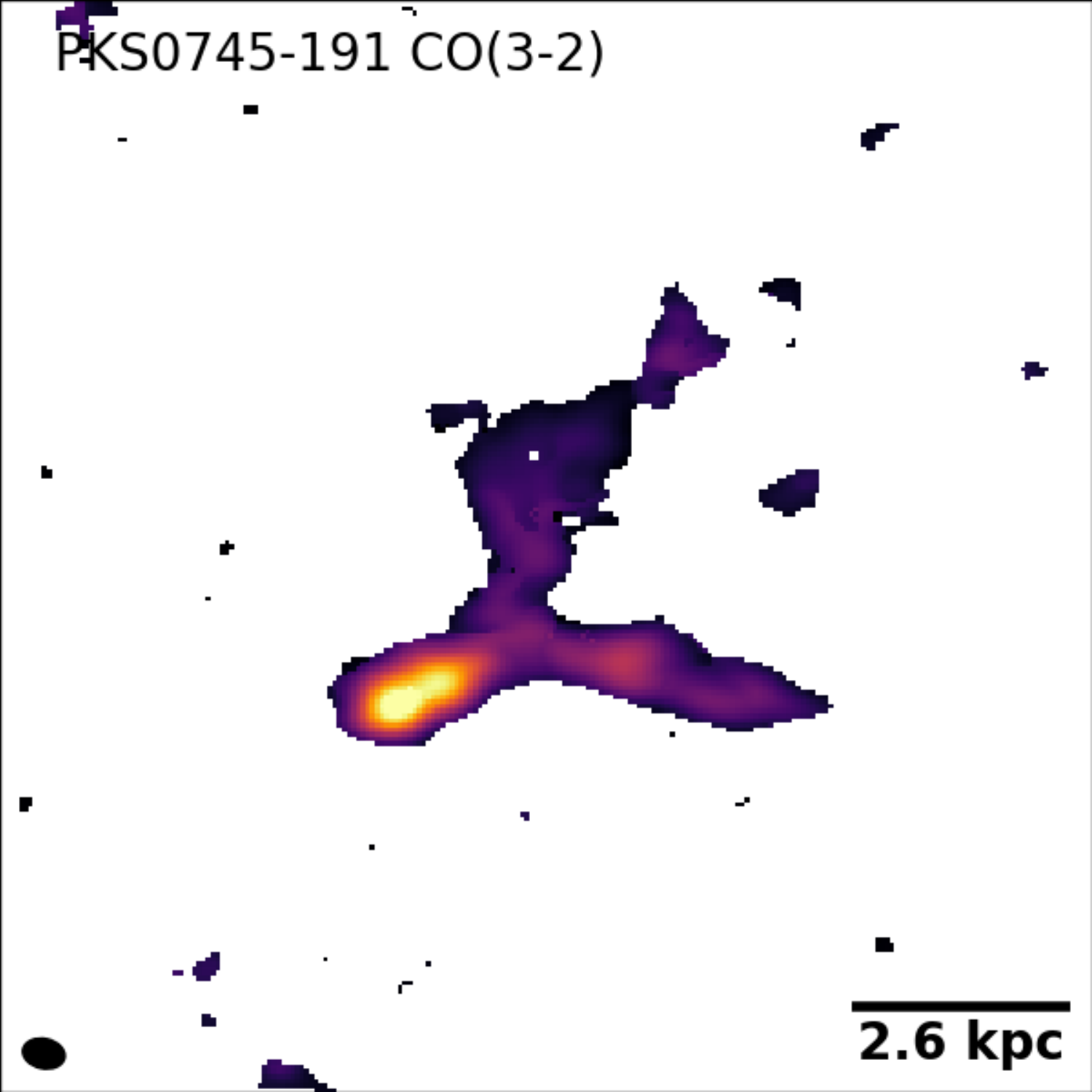}
	\caption{Comparison between total integrated flux maps for Zw3146 CO(1-0) 
		with PKS0745-191 CO(1-0) and CO(3-2).}
	\label{fig:pks_comp}
\end{figure*}

\subsection{Comparison to other BCGs}
\label{sec:PKS}

The majority of BCGs contain molecular gas distributions that do not resemble those in gas-rich spiral galaxies. Instead of rotationally-supported disks or rings, the cold gas in BCGs is generally filamentary \citep[e.g.][]{Olivares19, Russell19}. 
In a sample of 12 BCGs, only two have $\leq 10\%$ of their total molecular gas located in filaments \citep{Russell19}. The filaments in the remaining systems account for $30-100\%$ of their molecular gas supplies. These filaments can extend anywhere from several to several tens of kpc from the galactic center, often in the direction of X-ray cavities \citep[e.g.][]{mcn14, Russell16, Vantyghem16}. Their velocity gradients are often smooth, but are shallower than would be expected from free-fall. 

The three extensions in Zw~3146 CO(1-0) likely correspond to molecular filaments that are about $4\kpc$ in length. The outer portions of the filaments, identified by the boxes in Fig. \ref{fig:beamsub}, account for $40\%$ of the total flux. Radial filaments extending into the galactic center would contain an even larger fraction of the flux. Nevertheless, we are unable to rule out the presence of a molecular disk. The velocity structure, however, is inconsistent with the smooth gradient expected for a disk, so if a disk is present it must be unresolved. Similarly we do not observe the double-horned profile in the CO(1-0) spectrum that is typical of disks.

The system that compares most closely to the Zw~3146 CO(1-0) morphology is PKS0745-191, also at CO(1-0) \citep{Russell16}. Both exhibit a bright peak near the center of the gas distribution as well as three poorly-resolved radial extensions. However, the CO(3-2) observation of PKS0745-191 clearly resolves the three molecular filaments with no significant nuclear structure. The central peak at CO(1-0) results from the coarser resolution blending together the radial filaments. A comparison of Zw~3146 CO(1-0) with both PKS0745-191 CO observations is shown in Fig. \ref{fig:pks_comp}.

Subtracting the central, unresolved emission from the PKS0745-191 CO(1-0) data (see Section \ref{sec:kinematics} for the same analysis on Zw~3146), $52\%$ of the total flux can be attributed to an unresolved clump at the peak emission position. This is larger than the $22\%$ measured for Zw~3146 CO(1-0), while in PKS0745-191 CO(3-2) $<5\%$ of the emission is located at the vertex of its three filaments.
Without additional higher resolution observations of Zw~3146 we cannot conclusively determine its true molecular gas morphology. However, the similarities with PKS0745-191 suggest that it could share the morphology of three radial filaments with little central emission, observed with intermediate resolution.

\subsection{Origin of the Molecular Gas}

The massive ($5\e{10}\Msun$) reservoir of cold gas residing in the central 4~kpc of the Zw~3146 BCG places stringent requirements on its origin. Several tens of gas-rich spiral galaxies would need to be stripped to account for this gas supply. In cluster cores these are rare, as most central galaxies are red, gas-poor ellipticals \citep{Best07}. Sustained stellar mass loss will contribute to the molecular gas supply, but cannot account for the entire reservoir \citep{Sparks89, Voit11}. Instead, like most BCGs, the primary mechanism for forming the molecular gas is the condensation of the hot atmosphere.

Although the hot atmosphere contains ample fuel to produce the molecular gas reservoir, the region covered by molecular emission is dominated by molecular gas. Only $2.5\pm0.2\e{10}\Msun$ of hot gas is present within the central $10\kpc$, while the $5\e{10}\Msun$ of cold gas is located in the central $4\kpc$. Zw~3146 is not unique in this regard. In other systems observed by ALMA, the molecular gas mass is approximately equal to the hot gas mass when considering only the region that contains molecular emission \citep{Russell19}. Forming all of the molecular gas in this region would result in a significant inflow as the central hot atmosphere is depleted. Alternatively, the cold gas could be formed on larger scales in filaments that are too faint to be detected. The Ly-$\alpha$ emission, which could be coincident with this fainter molecular line emission, extends well beyond the inner 4~kpc in which the molecular gas is confined. Gas condensing out of the central $20\kpc$ would have $2\e{11}\Msun$ of hot gas to draw from.

In the stimulated feedback paradigm, molecular gas condenses from intracluster gas that has been entrained and uplifted by a radio bubble \citep{mcn16}. Zw~3146 harbours multiple X-ray cavities that have likely arisen from $2-3$ cycles of AGN activity. All three molecular extensions are projected behind an X-ray cavity. The SE and W extensions trail the two most prominent cavities -- A and B, respectively. The NE extension trails the most distant and poorly-defined cavity (F).

By Archimedes' principle, the maximum mass of gas that can be uplifted in the wake of a rising bubble is given by the mass it has displaced. This demand is alleviated by a factor of a few because the hot atmosphere is initially buoyantly neutral, making the gas easier to lift \citep{Su17}. The cavities that are associated with a molecular extension have each displaced several $10^{10}\Msun$ of hot gas (see Table \ref{tab:cavities}). This is several times larger than the mass of the molecular gas in the extensions (Table \ref{tab:spectra}). Combined, the cavities displace a total of $1.3\e{11}\Msun$ of hot gas. Thus these cavities could have initiated the formation of the entire $5\e{10}\Msun$ reservoir of cold gas.

Spectroscopic estimates of the mass deposition rate made using {\it Chandra} suggest that the hot atmosphere is condensing at a rate of $300\Msunpyr$ \citep{Egami06}. However, a recent spectral analysis conducted using archival data from the {\it XMM-Newton} Reflection Grating Spectrometer, following the prescription of \citet{Liu19}, measured $<50\Msunpyr$ of continuous radiative cooling from the bulk temperature ($4\keV$) down to $0.01\keV$ (Liu et al. in preparation). This includes $130\pm60\Msunpyr$ of gas cooling to $0.7\keV$, with $<78\Msunpyr$ cooling below that. This is closely matched to the star formation rate of $70\pm14\Msunpyr$. 
It would take $1\Gyr$ to produce the observed reservoir with condensation proceeding at $50\Msunpyr$. Condensation at this rate over the age of a single cavity would contribute a few $10^9\Msun$ of cold gas. The age of the AGN two most energetic AGN outbursts, determined by the buoyancy time of their respective cavities, are $\sim 4\e{7}\yr$ and $\sim 7\e{7}\yr$. This translates to the formation of $2-3.5\e{9}\Msun$ of molecular gas, $4-7\%$ of the total supply.

Stimulated feedback can therefore contribute significantly to the reservoir of molecular gas, but many cycles of AGN feedback would be required to account for the entire reservoir. 
Each cycle of AGN feedback can stimulate the formation of a few $10^9\Msun$ of molecular gas. This process must be sustained for 1~Gyr or longer in order to accumulate the cold gas mass that is observed. The consumption of cold gas in star formation complicates this picture, however, as the star formation rate is comparable to the mass deposition rate.

\section{Conclusions}

Our ALMA CO(1-0) observations of the Zw~3146 BCG have revealed a massive ($5\e{10}\Msun$) reservoir of molecular gas located within $4\kpc$ of the galactic center. The gas distribution is marginally resolved into three filaments. We have also identified six cavities in the {\it Chandra} X-ray image. Each molecular filament trails one of the X-ray cavities. 

The cavities have displaced a mass of intracluster gas several times larger than the molecular gas mass, suggesting that stimulated cooling -- where low entropy gas from the cluster core is uplifted by rising radio bubbles until it becomes thermally unstable and condenses -- may be responsible for the formation of the molecular gas. However, stimulated cooling would have difficulty accounting for the entire reservoir, as the cavity ages are more than ten times shorter than the 1~Gyr it would take to form $5\e{10}\Msun$ of cold gas at the observed condensation rate. The total gas supply is likely accumulated over many cycles of AGN feedback, with each cycle shaping its distribution.

While the molecular gas is centered within the BCG, the continuum source originating from the AGN is not. The continuum source is located $2.6\kpc$ away from the peak of the CO(1-0) emission, placing it at the outer extent of the molecular filaments. The displacement between the continuum source and the galactic center was likely caused by a merger. Possible tidal features within the BCG may be indicative of a merger remnant. The continuum source may correspond to the nucleus of this secondary galaxy. Alternatively, a gravitational wave recoil following a recent ($<10^7\yr$) black hole merger may account for this displacement. Alternative mechanisms -- such as a chance alignment with a background galaxy, the continuum corresponding to the lobe of a jet, or an asymmetric jet accelerating the SMBH of the BCG -- are unlikely.

\acknowledgements

We thank the anonymous referee for their helpful comments that improved this paper. ANV thanks Yjan Gordon and Cameron Lawlor-Forsyth for their support.
ANV, BRM, CPO, and SAB are supported by the Natural Sciences and Engineering Research Council of Canada.
This paper makes use of the ALMA data ADS/JAO.ALMA 2018.1.01056.S. ALMA is a partnership of the ESO (representing its member states), NSF (USA) and NINS (Japan), together with NRC (Canada), NSC and ASIAA (Taiwan), and KASI (Republic of Korea), in cooperation with the Republic of Chile. The Joint ALMA Observatory is operated by ESO, AUI/NRAO, and NAOJ.
This research made use of Astropy, a community-developed core Python package for Astronomy.
This research made use of APLpy, an open-source plotting package for Python hosted at http://aplpy.github.com.

\bibliographystyle{apj}
\bibliography{refs}

\end{document}